\documentclass{SciPost}

\hypersetup{
    colorlinks,
    linkcolor={red!50!black},
    citecolor={blue!50!black},
    urlcolor={blue!80!black}
}

\usepackage[bitstream-charter]{mathdesign}
\usepackage{tikz}
\usepackage{tabularx}

\DeclareSymbolFont{usualmathcal}{OMS}{cmsy}{m}{n}
\DeclareSymbolFontAlphabet{\mathcal}{usualmathcal}

\fancypagestyle{SPstyle}{
\fancyhf{}
\lhead{\colorbox{scipostblue}{\bf \color{white} ~SciPost Physics }}
\rhead{{\bf \color{scipostdeepblue} ~Submission }}

\fancyfoot[C]{\textbf{\thepage}}
}

\begin{document}

\pagestyle{SPstyle}

\begin{center}{\Large \textbf{\color{scipostdeepblue}{
%%%%%%%%%% TODO: Write your article's title here
Fusion approach for quantum integrable system associated with the $\mathfrak{gl}(1|1)$ Lie superalgebra\\
}}}\end{center}

\begin{center}
\textbf{
% Use (full) first name (+ middle name initials) + surname format.
% Separate subsequent authors by a comma, omit comma and use "and" for the last author.
% Mark the corresponding author(s) with a superscript symbol in this order
% \star, \dagger, \ddagger, \circ, \S, \P, \parallel, ...
Xiaotian Xu\textsuperscript{1,2,3,4},
Wuxiao Wen \textsuperscript{1},
Tao Yang\textsuperscript{1,2,3,4},
Xin Zhang\textsuperscript{5*},
Junpeng Cao\textsuperscript{2,5,6$^{\dagger}$
}}\end{center}

\begin{center}
{\bf 1} Institute of Modern Physics, Northwest University, Xi'an 710127, China
\\
{\bf 2} Peng Huanwu Center for Fundamental Theory, Xi'an 710127, China
\\
{\bf 3} Shaanxi Key Laboratory for Theoretical Physics Frontiers, Xi'an 710127, China \\
{\bf 4} Fundamental Discipline Research Center for  Quantum Science and Technology of Shaanxi Province, Xi'an 710127, China \\
{\bf 5} Beijing National Laboratory for Condensed Matter Physics, Institute of Physics, Chinese Academy of Sciences, Beijing 100190, China \\
{\bf 6} School of Physical Sciences, University of Chinese Academy of Sciences, Beijing 100049, China
%%%%%%%%%% END TODO: AFFILIATIONS
% Provide email address of corresponding author(s)
\\[\baselineskip]
$\star$ \href{mailto:email1}{\small xinzhang@iphy.ac.cn}\,,\quad
$\dagger$ \href{mailto:email2}{\small junpengcao@iphy.ac.cn}
\end{center}

\section*{\color{scipostdeepblue}{Abstract}}
\textbf{\boldmath{%
In this work we obtain the exact solution of quantum integrable system associated with the Lie superalgebra $\mathfrak{gl}(1|1)$, both for periodic
and for generic open boundary conditions.  By means of the fusion
technique we derive a closed set of operator identities among the fused
transfer matrices. These identities allow us to determine the complete
energy spectrum and the corresponding Bethe ansatz equations of the model.
Our approach furnishes a systematic framework for studying the
spectra of quantum integrable models based on Lie superalgebras, in
particular when the $U(1)$ symmetry is broken. The derivation of the Bethe states from the exact spectrum is also addressed.
}}

\vspace{\baselineskip}

%%%%%%%%%% BLOCK: Copyright information
% This block will be filled during the proof stage, and finilized just before publication.
% It exists here only as a placeholder, and should not be modified by authors.
\noindent\textcolor{white!90!black}{%
\fbox{\parbox{0.975\linewidth}{%
\textcolor{white!40!black}{\begin{tabular}{lr}%
  \begin{minipage}{0.6\textwidth}%
    {\small Copyright attribution to authors. \newline
    This work is a submission to SciPost Physics. \newline
    License information to appear upon publication. \newline
    Publication information to appear upon publication.}
  \end{minipage} & \begin{minipage}{0.4\textwidth}
    {\small Received Date \newline Accepted Date \newline Published Date}%
  \end{minipage}
\end{tabular}}
}}
}
%%%%%%%%%% BLOCK: Copyright information

% For convenience during refereeing we turn on line numbers:
\nolinenumbers

\vspace{10pt}
\noindent\rule{\textwidth}{1pt}
\tableofcontents
\noindent\rule{\textwidth}{1pt}
\vspace{10pt}
\section{Introduction}
	\setcounter{equation}{0}
	
	Quantum integrable models \cite{Baxter_book,Yang1967,Korepin1997} possess significant applications in quantum field theory, condensed matter physics and statistical physics, because the exact solutions of these models are crucial for understanding various strongly correlated effects
	and many-body physical mechanism.

Quantum integrable models associated with Lie superalgebras constitute a broad subclass of integrable systems \cite{FL2000}. Typical examples include
the $\mathrm{SU}(m|n)$ supersymmetric spin chains \cite{Li:2000dvq,Wlyang2006}, the Hubbard model \cite{Lieb68,Shastry86,Hubbard_book}, and the
supersymmetric $t$-$J$ model \cite{Wiegmann88,Essler92,Essler92_2}. These
models have applications in a variety of fields, such as disordered electronic
systems \cite{Mudry:1995zs}, critical phenomena in statistical mechanics
\cite{Read:2001pz}, and the AdS/CFT correspondence in string theory
\cite{Berkovits:1999im}.

The eigenvalue problem for this class of models can be tackled by either the coordinate Bethe ansatz (CBA) or the (nested) algebraic Bethe ansatz (ABA) \cite{Yang:2004pu,Gohmann:2001wh, Martins:1997hhu, Guan:1998ekr, Galleas:2007zz}. These approaches hinge on the existence of a reference (or pseudo-vacuum)
state. In the presence of a $U(1)$ symmetry, the reference state is readily
constructed. However, when the $U(1)$ charge is absent, the construction of the reference state becomes highly non-trivial and often impossible, severely limiting the applicability of the conventional Bethe ansatz techniques.

It has been recognized that a reference state is not indispensable for solving
the spectral problem. The off-diagonal Bethe ansatz (ODBA) \cite{ODBA_book} bypasses this
requirement by exploiting operator identities satisfied by the transfer matrix,
from which Baxter’s $T$-$Q$ relation can be constructed directly. Nevertheless, extending the ODBA to models based on
Lie superalgebras encounters several technical obstacles. A prominent example is
the Hubbard model: in order to obtain the full set of Bethe ansatz equations one still
has to perform a conventional coordinate Bethe ansatz or algebraic Bethe ansatz at the first nested level \cite{Martins:1997hhu,Li14}, which re-introduces the need for a suitable reference state.

Although significant progress has been made, solving integrable models associated with Lie superalgebras without invoking any reference state remains an open problem. In this work we address this challenge and propose a reference-state-free framework for these quantum integrable systems.

In the present study, we focus on $\mathfrak{gl}(1|1)$, one of the most elementary Lie superalgebras. In Ref. \cite{GAM2010} Grabowski and Frahm derived the spectrum of the $\mathfrak{gl}(1|1)$ superspin chain for diagonal and quasi-diagonal boundary conditions, imposing certain constraints.  Their analysis relied on the graded algebraic Bethe ansatz method, i.e., eigenstates were constructed by acting with creation operators on a properly chosen reference state. For generic non-diagonal boundary conditions, however, the construction of such a reference state becomes exceedingly difficult, rendering the conventional algebraic Bethe ansatz method inapplicable.

The purpose of the present paper is to extend the rigorous fusion techniques introduced in Refs. \cite{Karowski:1978ps, Kulish:1981gi, Kirillov:1987zz, Mezincescu:1991ke, Mezincescu:1991ag, Zhou:1995zy} to the graded case. Unlike the standard fusion procedure, we perform fusion along two branches. This yields a closed set of operator identities among the fused transfer matrices, from which the eigenvalue problem of the $\mathfrak{gl}(1|1)$ quantum integrable model are solved exactly.
With the exact spectrum in hand, we employ the separation of variables (SoV) approach \cite{Niccoli2012_1,Niccoli2012_2,Kitanine2015} to construct the Bethe state \cite{ODBA_book,zhang15,Zhang2014}.

The paper is organized as follows. In Section \ref{section2}, we study the  integrable model associated with $\mathfrak{gl}(1|1)$ under periodic boundary condition. The fusion procedure is employed to build the fused transfer matrices. We obtain a closed set of operator identities that determine their eigenvalues, which are parameterized by the well-known $T$-$Q$ relation. 
In Section \ref{section3}, we extend the fusion technique to the open boundary case. 
The eigenvalue problem of the system is solved through the operator identities regarding the fused transfer matrices. 
Section \ref{section4} presents the construction procedure for the Bethe states of the open $\mathfrak{gl}(1|1)$ integrable model.
We provide a conclusion in Section \ref{section5} 
	
\section{\texorpdfstring{$\mathfrak{gl}(1|1)$ integrable model with periodic boundary}{gl(1|1) integrable model with periodic boundary}}
\label{section2}
	
	\subsection{Integrability}
	
	Let $V$ be a $2$-dimensional $\mathbb{Z}_2$-graded linear space with a basis $\{|i\rangle|i = 1,2\}$, where the Grassmann parities are $p(1)=0$ and $p(2)=1$, which endows the 2-dimensional representation of the exceptional $\mathfrak{gl}(1|1)$ Lie superalgebra. The $R$-matrix $R(u) \in \mathrm{End}(V_1 \otimes_{s} V_2)$ of the
	supersymmetric $\mathfrak{gl}(1|1)$ model is \cite{Göhmann1998,GAM2010}
	\begin{equation}
		R_{1,2}(u)=\begin{pmatrix}u+\eta &  &  & \\  & u & \eta & \\  & \eta & u & \\  &  &  & u-\eta\end{pmatrix} ,\label{a}
	\end{equation}
	where $u$ is the spectral parameter and $\eta$ is the crossing parameter. Here and below we adopt the standard notations: for any matrix $A\in {\rm End}({V}\otimes_{s} {V})$, $A_{i,j}$ is a super embedding operator of $A$ in the graded tensor space, which acts as identity on the spaces except for the $i$-th and $j$-th ones. 
		
	The $R$-matrix (\ref{a}) possesses the following properties:
	\begin{align}
		\text{regularity} &:\quad R_{1,2}(0)=\eta P_{1,2}, \\
		\text{unitarity} &:\quad R_{1,2}(u) R_{2,1}(-u)=\rho_1(u)\times\mathbb{I},\quad\rho_1(u)=-(u-\eta)(u+\eta), \\
		\text{crossing-unitarity} &:\quad R_{1,2}^{st_{1}}(-u)R_{2,1}^{st_{1}}(u)= \rho_2(u)\times \mathbb{I},\quad\rho_2(u)=-u^2,
	\end{align}
	where $P_{1,2}$ is the super permutation operator. Here, $st_{i}$ is the partial super transposition  ($A_{i,j}^{st_{i}} = A_{j,i}(-1)^{p(i)[p(i)+p(j)]}$) \cite{Grabinski:2012qt}  and the super tensor product of two operators satisfies the rule $(A\otimes_{s}B)_{j l}^{i k}=(-1)^{[p(i)+p(j)]p(k)}A^{i}_{j}B^{k}_{l}$.
	The $R$-matrix (\ref{a}) satisfies the graded Yang-Baxter equation (GYBE) \cite{Kulish:1980ii, Kulish:1985bj, Göhmann1998}
	\begin{equation}
		R_{1,2}(u-v)R_{1,3}(u)R_{2,3}(v)=R_{2,3}(v)R_{1,3}(u)R_{1,2}(u-v).\label{b}
	\end{equation}
	We can construct the monodromy matrix $T(u)$ via the $R$-matrix (\ref{a}) as
	\begin{align}
		T_0(u)&=R_{0,1}(u-\theta_1)R_{0,2}(u-\theta_2)\cdots R_{0,N}(u-\theta_N)=\left(
		\begin{array}{cc}
		A(u) & B(u) \\
		C(u) & D(u)
		\end{array}
		\right).\label{c}
	\end{align}
	Here, $\{\theta_j \vert j=1,\ldots,N\}$ are inhomogeneous parameters, the subscript $0$ denotes the auxiliary space $V_0$, and the tensor product $V^{\otimes_s N}$ represents the physical (quantum) space, where $N$ is the number of lattice sites. 
	
	The monodromy matrix $T(u)$ satisfies the graded 
	RTT relation
	\begin{equation}
		R_{1,2}(u-v)T_1(u)T_2(v)=T_2(v)T_1(u)R_{1,2}(u-v),\label{d}
	\end{equation}
	 and can be expressed as a $2\times2$ matrix in the auxiliary space, whose entries are operators acting on $V^{\otimes_{s} N}$.
	
	Under periodic boundary condition, the transfer matrix of the system is defined as the super trace of the monodromy matrix in the auxiliary space
	\begin{equation}
	t_p(u)={\rm str}_0\{T_0(u)\}=\sum_{\alpha=1}^2(-1)^{p(\alpha)}[T_0(u)]_\alpha^\alpha.
	\end{equation}
	With the help of the RTT relation (\ref{d}), one can prove that the transfer matrices with different spectral parameters commute with each other, i.e., $[t_p(u), t_p(v)] = 0$, which guarantees the integrability of the system. 
	
	The Hamiltonian is given by the logarithmic derivative of the transfer matrix
	\begin{equation}
		\begin{aligned}\label{Ham}
			H_p&=\eta\left.\frac{\partial\ln t_p(u)}{\partial u}\right|_{u=0,\{\theta_j=0\}}=
			\sum_{j=1}^N{P_{j,j+1}}\\
			&=\sum_{j=1}^N\left(E_{j}^{11}E_{j+1}^{11}+E_{j}^{12}E_{j+1}^{21}+E_{j}^{21}E_{j+1}^{12}-E_{j}^{22}E_{j+1}^{22}\right),
		\end{aligned}  
	\end{equation}
	where $\{E_{k}^{ij}\}$ are generators of the superalgebra $\mathfrak{gl}(1|1)$, which act on the $k$-th quantum space, and the periodic boundary implies that $E_{N+1}^{ij}\equiv E_{1}^{ij}$. The generator ${E_{k}^{ij}}$ can be expressed in terms of the standard fermionic representation \[E_{k}^{11}=1-n_k,\qquad E_{k}^{12}=c_k,\qquad E_{k}^{21}=c^{\dagger}_k,\qquad E_{k}^{22}=n_k,\]	where ${c_j,\, c_j^\dagger}$ and $n_k$ denote the fermionic annihilation, creation, and particle number operators, respectively.
	Therefore, the Hamiltonian  \eqref{Ham} can be rewritten as \cite{GAM2010}
	\begin{equation}\label{t}
		H_p=\sum_{j=1}^N H_{j,j+1}=\sum_{j=1}^N \left(c_j^\dagger c_{j+1}+c_{j+1}^\dagger c_j-n_j-n_{j+1}\right) + N.	
	\end{equation}
	
	The Hamiltonian in Eq. (\ref{t}) describes a model of free fermions, which can be diagonalized directly. In this paper, we solve this model in the framework of Bethe ansatz.

		\subsection{\texorpdfstring{Fusion of the $R$-matrix}{Fusion of the R-matrix}}\label{Sec:FusionR}
	Fusion is a powerful and standard method for solving integrable models, particularly for those associated with high-rank Lie algebras.
	The $R$-matrix in integrable models degenerates into projection operators at some special points of spectral parameter $u$, which makes it possible to carry out the fused $R$-matrices and transfer matrices \cite{Karowski:1978ps,Kulish:1981gi, Kirillov:1987zz,Mezincescu:1991ke,Mezincescu:1991ag,Zhou:1995zy}.
	Within the conventional fusion approach, the procedure follows a single branch, as illustrated by the sequence
\begin{align*}
\mathfrak{t}(u)\rightarrow \mathfrak{t}^{(1)}(u)\rightarrow \mathfrak{t}^{(2)}(u)\cdots \rightarrow \mathfrak{t}^{(k)}(u).
 \end{align*}
The fusion procedure is considered closed when the highest-level fused transfer matrix $\mathfrak{t}^{(k)}(u)$ either becomes directly solvable \cite{Cao2014,Hao2016}  or coincides with a transfer matrix of lower level \cite{Li2019,Li2022}. In many ordinary (non-graded) models
this closure occurs after a finite number of fusion steps.

For the Lie superalgebra $\mathfrak{gl}(1|1)$ the situation is qualitatively different.  The fusion of the $R$-matrix along a single branch does not yield a closed form; instead, it requires a procedure carried out along two branches, as detailed in Sections \ref{Sec:first:fusion} and \ref{Sec:second:fusion}.

\subsubsection{First fusion
branch}\label{Sec:first:fusion}

	\paragraph{First-level fusion}
	At the point $u = \eta$, the $R$-matrix (\ref{a}) degenerates into a 2-dimensional supersymmetric projection operator $P_{1,2}^{(+)}$
	\begin{align}\label{pro1}
		R_{1,2}(\eta)=2\eta P_{1,2}^{(+)}.
	\end{align}
	Operator $P_{1,2}^{(+)}$ is defined by
	\begin{align}
	&P_{1,2}^{(+)}=\sum_{i=1}^2|\psi_i\rangle\langle \psi_i|,\qquad P_{1,2}^{(+)}=P_{2,1}^{(+)},\label{e}\\
	&|\psi_1\rangle=|1,1\rangle,\quad|\psi_2\rangle=\frac{1}{\sqrt{2}}(|1,2\rangle+|2,1\rangle),\label{A3}
	\end{align}
	with the parities
	\[p(\psi_1)=0,\quad p(\psi_2)=1,\]
	and projects the original 4-dimensional tensor space $V_1 \otimes_{s} V_2$ into a new 2-dimensional space spanned by $|\psi_1\rangle$ and $|\psi_2\rangle$. The projectors $P_{1,2}^{(+)}$ and $P_{2,1}^{(+)}$ can be obtained by exchanging two spaces $V_1$ and $V_2$, i.e., $|kl\rangle\rightarrow|lk\rangle$.

	 Using the projector $P_{2,1}^{(+)}$, we can construct the fused $R$-matrices
		\begin{eqnarray}
			&R_{\langle1,2\rangle,3}(u)=(u+\tfrac{1}{2}\eta)^{-1}P_{2,1}^{(+)}R_{1,3}(u-\tfrac{1}{2}\eta)R_{2,3}(u+\tfrac{1}{2}\eta)P_{2,1}^{(+)}\equiv R_{\bar{1},3}(u),\label{aaa}\\[3pt]
			&R_{3,\langle1,2\rangle}(u)=(u+\tfrac{1}{2}\eta)^{-1}P_{1,2}^{(+)}R_{3,1}(u-\tfrac{1}{2}\eta)R_{3,2}(u+\tfrac{1}{2}\eta)P_{1,2}^{(+)}\equiv R_{3,\bar{1}}(u),
	\end{eqnarray}
	where we denote the projected space by $V_{\bar{1}} = V_{\langle 1,2 \rangle} = V_{\langle 2,1 \rangle}$.
	
	The fused $R$-matrix ${R}_{\bar{1},n}(u)$ given by (\ref{aaa}) is a $4 \times 4$ matrix acting on the tensor space $V_{\bar{1}} \otimes_{s} V_n$. Its explicit form is
	\begin{equation}
			R_{\bar{1},n} (u)=
			\begin{pmatrix}
				u+\tfrac{3}{2}\eta & & \\[2pt]
				& u-\tfrac{1}{2}\eta & \sqrt{2} \eta \\[2pt]
				& \sqrt{2} \eta & u+\tfrac{1}{2}\eta \\[2pt]
				& & & u-\tfrac{3}{2}\eta
			\end{pmatrix}.
	\end{equation}
 %We see that the elements of fused $R$-matrices ${R}_{\bar{1},2}(u)$ and ${R}_{2,\bar{1}}(u)$ are the polynomials of $u$ with degree one.

	\paragraph{Second-level fusion}
	At the point of $u=-\tfrac{3}{2}\eta$,  the fused $R$-matrix defined in $R_{\bar{1},2} (u)$ (\ref{aaa}) degenerates into another projector
	\begin{equation}\label{pro2}
		R_{\bar{1},2}(-\tfrac{3}{2}\eta)=-3\eta \mathbb{P}_{\bar{1},2}^{(-)}.
	\end{equation}
	Here, $\mathbb{P}_{\bar{1},2}^{(-)}$ is a 2-dimensional supersymmetric projector
	\begin{equation}\label{fu2}
		\mathbb{P}_{\bar{1},2}^{(-)}=\sum_{i=1}^{2}|\phi_i\rangle\langle\phi_i|,
	\end{equation}
	where
	\begin{equation}\label{A8}
		|\phi_1\rangle=\frac{1}{\sqrt{3}}(\sqrt{2}|\psi_1\rangle\otimes_{s}|2\rangle-|\psi_2\rangle\otimes_{s}|1\rangle),\quad|\phi_2\rangle=|\psi_2\rangle\otimes_{s}|2\rangle.
	\end{equation}
	The basis vectors $|\phi_1\rangle$ and $|\phi_2\rangle$ have parities
	\[p(\phi_{1})=1,\quad p(\phi_{2})=0.\]
	We see that the operator $\mathbb{P}_{\bar{1},2}^{(-)}$ projects the original 4-dimensional tensor space $V_{\bar{1}} \otimes_{s} V_{2}$ into a new 2-dimensional space spanned by $|\phi_1\rangle$ and $|\phi_2\rangle$.
	
Performing the fusion procedure on $R_{\bar{1},n}(u)$ with the projector $\mathbb{P}_{\bar{1},2}^{(-)}$ yields the following second-level fused $R$-matrices
	\begin{eqnarray}
		&R_{\langle\bar{1},2\rangle,3}(u)=u^{-1}\mathbb{P}_{\bar{1},2}^{(-)}R_{2,3}(u+\eta)R_{\bar{1},3}(u-\tfrac{1}{2}\eta)\mathbb{P}_{\bar{1},2}^{(-)}\equiv R_{\tilde{1},3}(u),\label{f}\\[3pt]
		&R_{3,\langle\bar{1},2\rangle}(u)=u^{-1}\mathbb{P}_{2,\bar{1}}^{(-)}R_{3,2}(u+\eta)R_{3,\bar{1}}(u-\tfrac{1}{2}\eta)\mathbb{P}_{2,\bar{1}}^{(-)}\equiv R_{3,\tilde{1}}(u).
	\end{eqnarray}
	Here, the projected space is denoted by $V_{\tilde 1}=V_{\langle\bar 1,2\rangle}=V_{\langle2,\bar 1\rangle}$.
	The fused $R$-matrix $R_{\tilde{1},n}(u)$ is a $4\times 4$ matrix defined in the tensor space $V_{\tilde 1}\otimes_{s} V_{n}$ and reads
	\begin{align}
	 R_{\tilde{1},n}(u)=\begin{pmatrix}
		u + 2\eta & & \\
		& u - \eta & -\sqrt{3}\eta \\
		& -\sqrt{3}\eta & u + \eta \\
		& & & u - 2\eta
	\end{pmatrix}.\label{Fused_R2}
\end{align}

	\subsubsection{Second fusion branch}\label{Sec:second:fusion}
	It should be noted that the $R$-matrix of the $\mathfrak{gl}(1|1)$ algebra admits another distinct fusion branch beyond the one discussed above. 
	Given the similarity of the procedure, we only present the final results and detail the second fusion branch in Appendix \ref{APP:Fusion}.
	
 At the point $u=-\eta$, the $R$-matrix (\ref{a}) is proportional to a projector $P_{1,2}^{(-)}$
\begin{align}R_{1,2}(-\eta)=-2\eta{P}_{1,2}^{(-)}.
\end{align}
By performing the fusion with the projector ${P}_{2,1}^{(-)}$, we obtain the first-level fused $R$-matrices
		\begin{eqnarray}
			&R_{\langle1,2\rangle^{\prime},3}(u)=(u-\tfrac{1}{2}\eta)^{-1}{P}_{2,1}^{(-)}R_{1,3}(u+\tfrac{1}{2}\eta)R_{2,3}(u-\tfrac{1}{2}\eta){P}_{2,1}^{(-)}\equiv R_{\bar{1}^{\prime},3}(u) ,\label{bbb}\\[3pt]
			&R_{3,\langle1,2\rangle^{\prime}}(u)=(u-\tfrac{1}{2}\eta)^{-1}{P}_{1,2}^{(-)}R_{3,1}(u+\tfrac{1}{2}\eta)R_{3,2}(u-\tfrac{1}{2}\eta){P}_{1,2}^{(-)}\equiv R_{3,\bar{1}^{\prime}}(u),
	\end{eqnarray}
where the projected space is denoted as $V_{\bar 1^\prime}=V_{\langle1,2\rangle^\prime}=V_{\langle2,1\rangle^\prime}$.

At the point of $u=\tfrac{3}{2}\eta$, the fused matrix   $R_{\bar{1}^{\prime},2}(u)$ given by Eq. (\ref{bbb}) degenerates into a projector $\mathcal{P}_{\bar{1}^{\prime},2}^{(+)}$
\begin{align}
R_{\bar{1}^{\prime},2}(\tfrac{3}{2}\eta)=3\eta\mathcal{P}_{\bar{1}^{\prime},2}^{(+)}.
\end{align}
With the help of  $\mathcal{P}_{\bar{1}^{\prime},2}^{(+)}$, we obtain the following second-level fused $R$-matrices
\begin{eqnarray}
		&R_{\langle\bar{1}^{\prime},2\rangle,3}(u)=u^{-1}\mathcal{P}_{\bar{1}^{\prime},2}^{(+)}R_{2,3}(u-\eta)R_{\bar{1}^{\prime},3}(u+\tfrac{1}{2}\eta)\mathcal{P}_{\bar{1}^{\prime},2}^{(+)}\equiv R_{\tilde{1}^{\prime},3}(u),\label{g}\\[3pt]
		&R_{3,\langle\bar{1}^{\prime},2\rangle}(u)=u^{-1}\mathcal{P}_{2,\bar{1}^{\prime}}^{(+)}R_{3,2}(u-\eta)R_{3,\bar{1}^{\prime}}(u+\tfrac{1}{2}\eta)\mathcal{P}_{2,\bar{1}^{\prime}}^{(+)}\equiv R_{3,\tilde{1}^{\prime}}(u),
	\end{eqnarray}
	where we denote the projected space as
	$V_{\tilde 1^\prime}=V_{\langle \bar 1^\prime,2\rangle}=V_{\langle 2,\bar 1^\prime \rangle}$.
	
\subsubsection{Closure of the fusion}

By a direct analysis, we find that $R_{\tilde{1},2}(u)$ given by (\ref{f}) and $R_{\tilde{1}^\prime,2}(u)$ given by (\ref{g}) are identical
	\begin{equation}
		R_{\tilde{1},2}(u)=R_{\tilde{1}^\prime,2}(u).\label{m}
	\end{equation}
	
\begin{figure}
\centering
\begin{tikzpicture}[node distance=2cm, scale=0.4]
\tikzstyle{startstop} = [rectangle, rounded corners, minimum width=1cm, minimum height=1cm,text centered, draw=black, fill=red!40]
\tikzstyle{io} = [rectangle, rounded corners, minimum width=1cm, minimum height=1cm, text centered, draw=black, fill=blue!40]
\tikzstyle{process} = [rectangle,rounded corners, minimum width=1cm, minimum height=1cm, text centered, draw=black, fill=orange!40]
\tikzstyle{decision} = [rectangle, rounded corners,, minimum width=1cm, minimum height=1cm, text centered, draw=black, fill=green!40]
\tikzstyle{arrow} = [thick,->,>=stealth]
\tikzstyle{darrow} = [thick,-,>=stealth]

\node (start1) [startstop,align=center] at (4,-0.5) {$R_{1,2}(u)$};

\node (in1) [io,align=center] at (1,-7) {$R_{\bar 1,2}(u)$};
\node[align=center] at (1,-3) {$P^{(+)}_{2,1}$};
\node (in2) [io,align=center] at (7,-7) {$R_{\bar 1',2}(u)$};
\node[align=center] at (7,-3) {$P^{(-)}_{2,1}$};

\node (in5) [process, align=center] at (1,-13){$R_{\tilde 1,2}(u)$};
\node[align=center] at (-0.5,-10) {$\mathbb{P}_{\bar{1},2}^{(-)}$};
\node[align=center] at (8.5,-10) {$\mathcal{P}_{\bar{1}^{\prime},2}^{(+)}$};

\node (in6) [process, align=center] at (7,-13) {$R_{\tilde 1',2}(u)$};
\node[align=center] at (4,-13) {$=$};

% 绘制箭头
\draw [arrow,blue] (3.7,-1.8)  -- (3.7,-4) -| (in1);
\draw [arrow,red] (4.3,-1.8)  -- (4.3,-4) -| (in2);
\draw [arrow,blue] (in1) -- (in5);
\draw [arrow,red] (in2) -- (in6);
%\draw [darrow] (in5) -- (in6);
\end{tikzpicture}
    \caption{The fusion procedure of  $R$-matrix.}
    \label{fig1}
\end{figure}
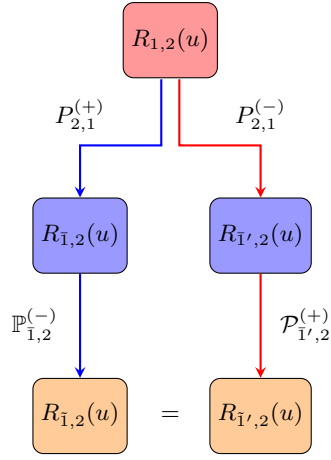

We perform fusion along two branches and connect the resulting fused $R$-matrices at the second fusion level. This connection thereby closes the fusion procedure, a mechanism quite different from the standard one. 
The fusion procedure of the $R$-matrix is briefly illustrated in Fig. \ref{fig1}.
	
	\subsection{Fused transfer matrices}
		
	The fused $R$-matrices satisfy the following graded Yang-Baxter equations
	\begin{equation}
	R_{\alpha,\beta}(u-v)R_{\alpha,\gamma}(u)R_{\beta,\gamma}(v)=R_{\beta,\gamma}(v)R_{\alpha,\gamma}(u)R_{\alpha,\beta}(u-v),\label{a1a}
	\end{equation}
where the indices $\alpha, \beta, \gamma$ may label either the original spaces or the projected spaces.

	Using the fused $R$-matrices defined in (\ref{aaa}), (\ref{f}), (\ref{bbb}), and (\ref{g}), we define the fused monodromy matrices
	\begin{equation}\label{FM1}
		T_{\alpha}(u) = R_{\alpha,1}(u - \theta_1) R_{\alpha,2}(u - \theta_2) \cdots R_{\alpha,N}(u - \theta_N),
	\end{equation}
	where the subscript $\alpha \in \{\bar{0}, \tilde{0}, \bar{0}^\prime, \tilde{0}^\prime\}$ refers to the  fused auxiliary spaces.  Here, $\bar{0}$ and $\tilde{0}$ correspond to the first-level and second-level of the first fusion branch respectively; whereas $\bar{0}'$ and $\tilde{0}'$ correspond to the first-level and second-level of the second fusion branch respectively.
	All the fused monodromy matrices in Eq. \eqref{FM1} satisfy the graded RTT relations
	\begin{equation}
		R_{\alpha,\beta}(u - v)\, T_{\alpha}(u)\, T_{\beta}(v)
		= T_{\beta}(v)\, T_{\alpha}(u)\, R_{\alpha,\beta}(u - v).\label{RTT-1}
	\end{equation}
	%{\color{blue} where $(\alpha$, $\beta)$ take the values from the set $\{ (\bar{1}, 2),\,\, (\bar{1}^\prime, 2),\,\, (\tilde{1}, 2),\,\,(\tilde{1}^{\prime}, 2)\}$.}
		
	The super traces of the fused monodromy matrices in the auxiliary spaces give the corresponding fused transfer matrices
	\begin{align}
	\begin{aligned}\label{fused;transfer;matrix}
		&t_{p}^{(1)}(u)={\rm str}_{\bar{0}}\{T_{\bar{0}}(u)\},\qquad t_p^{(2)}(u)={\rm str}_{\bar{0}^{\prime}}\{T_{\bar{0}^{\prime}}(u)\},\\ &\tilde{t}_p^{(1)}(u)={\rm str}_{\tilde{0}}\{T_{\tilde{0}}(u)\},
		\qquad\tilde{t}_p^{(2)}(u)={\rm str}_{\tilde{0}^{\prime}}\{T_{\tilde{0}^{\prime}}(u)\}.
	\end{aligned}
	\end{align}
	From Eq. (\ref{m}), we conclude that the fused transfer matrices $\tilde{t}_{p}^{(1)}(u)$ and $\tilde{t}_{p}^{(2)}(u)$ are identical, we therefore denote them collectively as $\tilde{t}_{p}(u)$:
	\begin{equation}
	\tilde{t}_{p}(u)=\tilde{t}_{p}^{(1)}(u)=\tilde{t}_{p}^{(2)}(u).\label{o}
	\end{equation}
	The graded RTT relations in (\ref{RTT-1}) imply that the transfer matrices $t_p(u)$, $t_{p}^{(1)}(u)$, $t_{p}^{(2)}(u)$ and $\tilde{t}_{p}(u)$ commute with each other, namely,
 \begin{equation}\label{dy1}
	\begin{aligned}	&[t_p(u),\,t_p^{(1)}(v)]=[t_p(u),\,t_p^{(2)}(v)]=[t_p^{(1)}(u),\,t_p^{(2)}(v)]=0,\\[3pt]
	&[\tilde{t}_{p}(u),\,t_p(v)]=[\tilde{t}_{p}(u),\,t_p^{(1)}(v)]=[\tilde{t}_{p}(u),\,t_p^{(2)}(v)]=0.
	\end{aligned}
	\end{equation}

    \subsection{Operator identities}
	The definitions of the fused $R$-matrices in (\ref{aaa}), (\ref{f}), (\ref{bbb}), and (\ref{g}) directly yield the following relations for the fused monodromy matrices
		\begin{equation}
				\begin{aligned}
					&P_{2,1}^{(+)}T_1(u)T_2(u+\eta)P_{2,1}^{(+)}=a(u+\eta)T_{\bar{1}}(u+\tfrac{1}{2}\eta),\\ &{P}_{2,1}^{(-)}T_1(u)T_2(u-\eta){P}_{2,1}^{(-)}=a(u-\eta)T_{\bar{1}^{\prime}}(u-\tfrac{1}{2}\eta),\\ &\mathbb{P}_{\bar{1},2}^{(-)}T_2(u+\eta)T_{\bar{1}}(u-\tfrac{1}{2}\eta)\mathbb{P}_{\bar{1},2}^{(-)}=a(u)T_{\tilde{1}}(u),\\ &\mathcal{P}_{\bar{1}^{\prime},2}^{(+)}T_2(u-\eta)T_{\bar{1}^{\prime}}(u+\tfrac{1}{2}\eta)\mathcal{P}_{\bar{1}^{\prime},2}^{(+)}=a(u)T_{\tilde{1}^{\prime}}(u),\end{aligned}\label{i}
		\end{equation}	
where 
\begin{align}
    a(u)=\prod_{j=1}^N(u-\theta_j).
\end{align}
From the graded RTT relations (\ref{RTT-1}) at specific points, together with the properties of the projectors, we derive
	\begin{equation}
			\begin{aligned}
				& T_1(\theta_j)T_2(\theta_j+\eta)=P_{2,1}^{(+)}T_1(\theta_j)T_2(\theta_j+\eta),\\
				& T_1(\theta_j)T_2(\theta_j-\eta)={P}_{2,{1}}^{(-)}T_1(\theta_j)T_2(\theta_j-\eta),\\
				& T_2(\theta_j)T_{\bar{1}}(\theta_j-\tfrac{3}{2}\eta)=\mathbb{P}_{\bar{1},2}^{(-)}T_2(\theta_j)T_{\bar{1}}(\theta_j-\tfrac{3}{2}\eta),\\
				& T_2(\theta_j)T_{\bar{1}^{\prime}}(\theta_j+\tfrac{3}{2}\eta)=\mathcal{P}_{\bar{1}^{\prime},2}^{(+)}T_2(\theta_j)T_{\bar{1}^{\prime}}(\theta_j+\tfrac{3}{2}\eta),
			\end{aligned}\label{i-11}
	\end{equation}
	where $j=1,\ldots,N$.
Taking the super trace of Eq. (\ref{i}) over the auxiliary space and using Eq. (\ref{i-11}), we obtain the operator product identities
\begin{align}
	\begin{aligned}
	&t_p(\theta_j)t_p(\theta_j+\eta)=a(\theta_j+\eta)t_p^{(1)}(\theta_j+\tfrac{1}{2}\eta),\\[3pt]
	&t_p(\theta_j-\eta)t_p(\theta_j)=a(\theta_j-\eta)t_p^{(2)}(\theta_j-\tfrac{1}{2}\eta),\\[3pt]
 	&t_p^{(1)}(\theta_j-\tfrac{3}{2}\eta)t_p(\theta_j)=a(\theta_j-\eta)\tilde{t}_p(\theta_j-\eta),\\[3pt]
	&t_p^{(2)}(\theta_j+\tfrac{3}{2}\eta)t_p(\theta_j)=a(\theta_j+\eta)\tilde{t}_p(\theta_j+\eta),
	\end{aligned}\label{p}
	\end{align}
	with $j=1,\ldots,N$.
	
\begin{figure}[htbp]
\centering
\begin{tikzpicture}[node distance=2cm, scale=0.4]
\tikzstyle{startstop} = [rectangle, rounded corners, minimum width=1cm, minimum height=1cm,text centered, draw=black, fill=red!40]
\tikzstyle{io} = [rectangle, rounded corners, minimum width=1cm, minimum height=1cm, text centered, draw=black, fill=blue!40]
\tikzstyle{process} = [rectangle,rounded corners, minimum width=1cm, minimum height=1cm, text centered, draw=black, fill=orange!40]
\tikzstyle{decision} = [rectangle, rounded corners,, minimum width=1cm, minimum height=1cm, text centered, draw=black, fill=green!40]
\tikzstyle{arrow} = [thick,->,>=stealth]
\tikzstyle{darrow} = [thick,<->,>=stealth]

\node (start1) [startstop,align=center] at (1,0) {$t_p(u_1)$};
\node (start2) [startstop, align=center] at (7,0) {$t_p(u_2)$};

\node at (4,0) {$\times $};
\node at (-2,-8) {$\times $};
\node at (10,-8) {$\times $};

\node (in1) [io,align=center] at (1,-8) {$t_p^{(1)}(u_3)$};
\node (in2) [io,align=center] at (7,-8){$t_p^{(2)}(u_4)$};

\node (in3) [startstop,align=center] at (-5,-8) {$t_p(u_5)$};
\node (in4) [startstop,align=center] at (13,-8){$t_p(u_6)$};
\node (in5) [process, align=center] at (4,-16){$\tilde{t}_p(u_7)$};

\draw [arrow, blue] (3.7,-1)  -- (3.7,-4) -| (in1);
\draw [arrow,red] (4.3,-1)  -- (4.3,-4) -| (in2);
\draw [arrow,blue] (-2,-9) -- (-2,-12)  -- (3.7, -12) -- (3.7,-14.7);
\draw [arrow,red] (10,-9) -- (10,-12)  -- (4.3, -12) -- (4.3,-14.7);

\end{tikzpicture}
\caption{Schematic diagram of the transfer matrix fusion procedure. The blue and red lines represent the fist and second fusion branches respectively. The spectral parameter $u_j$ must be set to a specific value at each step, as shown in Eq. \eqref{p}.}
\label{fig2}
\end{figure}
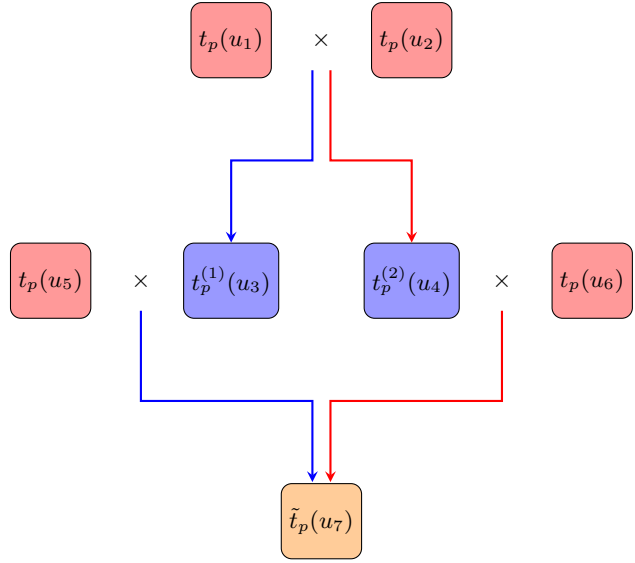

Figure \ref{fig2} shows a schematic of the transfer matrix fusion. Unlike the conventional approach, the procedure follows two fusion branches:
\begin{align}
(1): t_p(u)\rightarrow t_p^{(1)}(u)\rightarrow \tilde t_p^{(1)}(u), \quad
(2): t_p(u)\rightarrow t_p^{(2)}(u)\rightarrow \tilde t_p^{(2)}(u).
\end{align}
The fusion procedure is closed by the identity $\tilde t_p^{(1)}(u) = \tilde t_p^{(2)}(u)$. This suggests a novel strategy for solving integrable models associated with Lie superalgebra: building multiple fusion branches and connecting them to achieve a closed system.

	\subsection{\texorpdfstring{$T$-$Q$ relation}{TQ relation}}
	Let $\Lambda_p(u)$, $\Lambda_p^{(1)}(u)$, $\Lambda_p^{(2)}(u)$ and $\tilde{\Lambda}_p(u)$ denote the eigenvalues of the transfer matrices $t_p(u)$, $t_p^{(1)}(u)$, $t_p^{(2)}(u)$ and $\tilde{t}_p(u)$, respectively.
	As the fused transfer matrices mutually commute, the operator product identities in (\ref{p}) directly lead to the following functional relations
	\begin{equation}
	\begin{aligned}\label{r}			&\Lambda_{p}(\theta_j)\Lambda_{p}(\theta_j+\eta)=a(\theta_j+\eta)\Lambda_{p}^{(1)}(\theta_j+\tfrac{1}{2}\eta),\\[3pt]
	&\Lambda_{p}(\theta_j-\eta)\Lambda_{p}(\theta_j)=a(\theta_j-\eta)\Lambda_{p}^{(2)}(\theta_j-\tfrac{1}{2}\eta),\\[3pt]
	&\Lambda_{p}^{(1)}(\theta_j-\tfrac{3}{2}\eta)\Lambda_{p}(\theta_j)=a(\theta_j-\eta)\tilde{\Lambda}_p(\theta_j-\eta),\\[3pt]
	&\Lambda_{p}^{(2)}(\theta_j+\tfrac{3}{2}\eta)\Lambda_{p}(\theta_j)=a(\theta_j+\eta)\tilde{\Lambda}_p(\theta_j+\eta),
	\end{aligned}
	\end{equation}
	where $j=1,\ldots,N$.
	 Since $\Lambda_p(u)$, $\Lambda_p^{(1)}(u)$, $\Lambda_p^{(2)}(u)$, and $\tilde{\Lambda}_p(u)$ are degree-$(N-1)$ polynomials in $u$, the $4N$ constraints in Eq. (\ref{r}) completely determine these functions.
	 
	We can parameterize the eigenvalues $\Lambda_p(u)$, $\Lambda_p^{(1)}(u)$, $\Lambda_p^{(2)}(u)$ and $\tilde\Lambda_p(u)$ in terms of the following $T$-$Q$ relations
		\begin{equation}
		\begin{aligned}\label{TQ:periodic}
			&\Lambda_p(u)=[a(u)-a(u-\eta)]\frac{Q(u+\eta)}{Q(u)},\\
			&\Lambda_{p}^{(1)}(u)=[a(u-\tfrac{1}{2}\eta)-a(u-\tfrac{3}{2}\eta)]\frac{Q(u+\tfrac{3}{2}\eta)}{Q(u-\tfrac{1}{2}\eta)},\\
			&\Lambda_{p}^{(2)}(u)=[a(u-\tfrac{3}{2}\eta)-a(u-\tfrac{1}{2}\eta)]\frac{Q(u+\tfrac{3}{2}\eta)}{Q(u-\tfrac{1}{2}\eta)},\\
			&\tilde{\Lambda}_{p}(u)=[a(u-2\eta)-a(u-\eta)]\frac{Q(u+2\eta)}{Q(u-\eta)},
		\end{aligned}
	\end{equation}
	where
	\begin{equation}
		 Q(u)=\prod_{k=1}^M(u-\mu_k),
	\end{equation}
	and $M$ is the number of Bethe roots $\{\mu_{k}^{}\}$ and ranges from 0 to $N$. The analyticity of  $\Lambda_p(u)$, $\Lambda_p^{(1)}(u)$, $\Lambda_p^{(2)}(u)$ and $\tilde\Lambda_p(u)$ requires that the Bethe roots $\{\mu_{k}\}$ must satisfy the Bethe ansatz equations (BAEs)
	\begin{equation}\label{BAEs1}
		\prod_{j=1}^{N}\frac{\mu_{k}-\theta_{j}-\eta}{\mu_{k}-\theta_{j}}=1,\quad k=1,\ldots,M.
	\end{equation}

	The eigenvalue of the Hamiltonian (\ref{t}) can be given by the Bethe roots as follows
	\begin{equation}\label{en1}
	E_p=\eta\left.\frac{\partial\ln \Lambda_p(u)}{\partial u}\right|_{u=0,\{\theta_j=0\}}=	\sum_{k=1}^{M}\frac{\eta^2}{(\eta-\mu_{k})\mu_k}-N.
	\end{equation}
	
	Numerical results for the $N=3$ and $N=4$ cases are presented in Tables \ref{tab1} and \ref{tab2} respectively.  It can be seen that the eigenvalue $E_p$ derived from the Bethe roots coincides with that from the direct diagonalization of the Hamiltonian (\ref{t}).
	
	\begin{table}[htbp]
		\caption{Numeric results of Bethe roots $\{\mu_k\}$ and eigenvalues  of the Hamiltonian (\ref{t}). Here, $N=3$, $\eta=1$ and $\{\theta_j=0\}$.}
		\centering
		\vspace{\baselineskip}
		\renewcommand{\arraystretch}{1.4}
		\begin{tabularx}{0.45\linewidth}{|>{\centering\arraybackslash}X|>{\centering\arraybackslash}X|>{\centering\arraybackslash}X|>{\centering\arraybackslash}X|}
			\hline 
			$\mu_1$ & $\mu_2$ &  $\mu_3$& $E_p$ \\
			\hline 
			--    &  --  & -- & $-$3  \\
			$\infty$    & --  & -- & $-$3 \\
			$\frac{3-\mathrm{i} \sqrt{3}}{6}$  & --  & -- & \makebox[1.2em][r]{0} \\
			$\frac{3+\mathrm{i} \sqrt{3}}{6}$  & --  & -- & \makebox[1.2em][r]{0} \\
			$\frac{3-\mathrm{i} \sqrt{3}}{6}$  & $\infty$  & -- & \makebox[1.2em][r]{0} \\
			$\frac{3+\mathrm{i} \sqrt{3}}{6}$  & $\infty$  & -- & \makebox[1.2em][r]{0} \\
			$\frac{3+\mathrm{i} \sqrt{3}}{6}$  & $\frac{3-\mathrm{i} \sqrt{3}}{6}$ & -- & \makebox[1.2em][r]{3} \\
			$\frac{3+\mathrm{i} \sqrt{3}}{6}$ & $\frac{3-\mathrm{i} \sqrt{3}}{6}$ & $\infty$ & \makebox[1.2em][r]{3} \\
			\hline 
		\end{tabularx}
		\label{tab1}
	\end{table}
	
	\begin{table}[htbp]
		\centering
		\caption{Numeric results of Bethe roots $\{\mu_k\}$ and eigenvalues of the Hamiltonian (\ref{t}). Here, $N=4$, $\eta=1$ and $\{\theta_j=0\}$.}.
		\renewcommand{\arraystretch}{1.5}
		\vspace{\baselineskip}
		\begin{tabularx}{0.45\linewidth}{|>{\centering\arraybackslash}X|>{\centering\arraybackslash}X|>{\centering\arraybackslash}X|>{\centering\arraybackslash}X|>{\centering\arraybackslash}X|}
			\hline
			$\mu_1$ & $\mu_2$ &  $\mu_3$& $\mu_4$& $E_p$ \\
			\hline
			-- & -- & -- & -- & $-$4  \\
			$\infty $ & -- & -- & -- & $-$4  \\
			$\frac{1+\mathrm{i}}{2}$ & -- & -- & -- & $-$2 \\
			$\frac{1-\mathrm{i}}{2}$ & -- & -- & -- & $-$2 \\
			$\frac{1}{2}$ & -- & -- & -- &  \makebox[1.2em][r]{0}  \\
			$\infty $ & $ \frac{1+\mathrm{i}}{2} $ & -- & -- & $-$2 \\
			$\infty $ & $ \frac{1-\mathrm{i}}{2} $ & -- & -- & $-$2 \\
			$ \infty $ & $ \frac{1}{2} $ & -- & -- &  \makebox[1.2em][r]{0}  \\
			\hline
		\end{tabularx}
		\quad
		\begin{tabularx}{0.45\linewidth}{|>{\centering\arraybackslash}X|>{\centering\arraybackslash}X|>{\centering\arraybackslash}X|>{\centering\arraybackslash}X|>{\centering\arraybackslash}X|}
			\hline
			$\mu_1 $ &$\mu_2 $ & $ \mu_3 $ & $ \mu_4 $ & $ E_p$ \\
			\hline
			$\frac{1+\mathrm{i}}{2} $ & $ \frac{1-\mathrm{i}}{2} $ & -- & -- & 0  \\
			$ \frac{1+\mathrm{i}}{2}$ & $ \frac{1}{2} $ & -- & -- &  2  \\
			$ \frac{1-\mathrm{i}}{2} $ & $ \frac{1}{2} $ & -- & -- & 2  \\
			$ \infty $ &$ \frac{1+\mathrm{i}}{2} $ & $ \frac{1-\mathrm{i}}{2} $ & -- &  0  \\
			$ \infty $ & $ \frac{1-\mathrm{i}}{2} $ & $ \frac{1}{2} $ & -- & 2 \\
			$ \infty $ & $ \frac{1+\mathrm{i}}{2} $ & $ \frac{1}{2}$ & -- & 2 \\
			$ \frac{1+\mathrm{i}}{2} $ & $ \frac{1-\mathrm{i}}{2} $ & $ \frac{1}{2} $ & -- & 4 \\
			$ \infty $ & $ \frac{1+\mathrm{i}}{2} $ & $\frac{1-\mathrm{i}}{2} $ & $ \frac{1}{2} $ & 4 \\
			\hline
		\end{tabularx}
		\label{tab2}
	\end{table}
	
Under periodic boundary conditions, the $\mathfrak{gl}(1|1)$ integrable model possesses $U(1)$ symmetry, and common eigenstates of the transfer matrix and the Hamiltonian can be constructed as \cite{Korepin1997}
\begin{align}
|\mu_1,\ldots,\mu_M\rangle= \prod_{k=1}^MB(\mu_1)|0\rangle_1\otimes_s|0\rangle_2\cdots\otimes_s |0\rangle_N,\label{BS:pbc}
\end{align}
where $\{\mu_1,\ldots,\mu_M\}$ satisfy BAEs (\ref{BAEs1}) and $|0\rangle$ is the vacuum of the fermion.

		\section{\texorpdfstring{$\mathfrak{gl}(1|1)$ integrable model with open boundary}{gl(1|1) integrable model with open boundary}}\label{section3}
	
	\subsection{Integrability}
	In this section, we consider the $\mathfrak{gl}(1|1)$ integrable model under open boundary condition. Let us introduce the $K$-matrices $K^{-}(u)$ and $K^{+}(u)$. The matrix $K^{-}(u)$ satisfies the graded reflection equation (RE) \cite{Bracken:1997zww,Zhou:1998joy}
	\begin{equation}\label{b1}
		R _{1,2}(u-v){K^{-}_{  1}}(u)R _{2,1}(u+v) {K^{-}_{2}}(v)=
		{K^{-}_{2}}(v)R _{1,2}(u+v){K^{-}_{1}}(u)R _{2,1}(u-v),
	\end{equation}
	while $K^{+}(u)$ satisfies the graded dual reflection equation
	\begin{eqnarray}\label{b2}
		R_{1,2}(v-u)K_1^+(u)R_{2,1}(-u-v)K_2^+(v)=K_2^+(v)R_{1,2}(-u-v)K_1^+(u)R_{2,1}(v-u).
	\end{eqnarray}
	The generic solutions for the $K^{\pm}(u)$ are \cite{GAM2010}
	\begin{equation}\label{b3}
		K^{\pm}(u)=\mathbb{I}+u\begin{pmatrix}a_\pm&b_\pm\boldsymbol{\mathcal{E}}\\\\f_\pm\boldsymbol{\mathcal{E}}^\sharp&-a_\pm \end{pmatrix},
	\end{equation}
	where $a_\pm$, $b_\pm$ and $f_\pm$ are complex boundary parameters, ${\mathcal{E}}$ is the sole generator of complex Grassmann algebra $CG_1$, and ${\mathcal{E}}^\sharp$  is the adjoint of ${\mathcal{E}}$, i.e., ${\mathcal{E}}^\sharp=-\mathrm{i}{\mathcal{E}}$. Further details about Grassmann numbers ${\mathcal{E}}$ and  ${\mathcal{E}}^\sharp$ are provided in Appendix \ref{APP;Grassmann}.

	We should note that for the supersymmetric $\mathfrak{gl}(1|1)$ model, the $K$-matrices must be diagonal if they do not possess an additional internal space, i.e.,  all the elements are c-numbers. This implies that in a conventional Boson-Fermion mixture, bosons cannot transform into fermions upon boundary reflection. In contrast, the introduction of Grassmann numbers in Eq. \eqref{b3} allows for non-vanishing off-diagonal matrix elements.
	
	We notice that $[K^-(u),\,K^+(v)]$ $\neq 0$, which means that they cannot be diagonalized simultaneously. In this case, it is quite hard to obtain the eigenvalues via the conventional Bethe ansatz methods due to the lack of a proper reference state. 
	
	The transfer matrix $t(u)$ is constructed as
	\begin{equation}
		t(u)= {\rm str}_0 \{K_0^{ +}(u)T_0 (u) K^{ -}_0(u)\hat{T}_0 (u)\},\label{b5}
	\end{equation}
	where $\hat{T}(u)$ is the reflecting monodromy matrix
	\begin{eqnarray}
		\hat{T}_0 (u)=R_{N,0}(u+\theta_N)\cdots R_{2,0}(u+\theta_{2}) R_{1,0}(u+\theta_1).\label{Tt11}
	\end{eqnarray}

	By using the graded Yang-Baxter relations (\ref{b}) and reflection equations (\ref{b1})-(\ref{b2}) repeatedly,
	we can prove that the transfer matrices with different spectral parameters commute with each other. Therefore, $t(u)$ serves
	as the generating function of conserved quantities.
	The Hamiltonian is generated from the second-order derivative of the transfer matrix \cite{GAM2010}
	\begin{equation}
		\begin {aligned}
		H&=\frac{1}{8\eta^{N}(1+a_{+}\eta)}  \left.\frac{\partial ^2 t(u)}{\partial u^2}\right|_{u=0,\{\theta_j=0\}}\\&=\sum_{j=1}^{N-1}H_{j,j+1}+\frac{\eta^{N-1}}{2}\left[a_{-}-2a_{-}n_{1}+b_{-}\mathcal{E}c_{1}+f_{-}\mathcal{E}^{\sharp}c_{1}^{\dagger}\right]\\
		&\hspace{5mm}+\frac{\eta^{N-1}}{2(1+a_{+}\eta)}\left[a_{+}-2a_{+}n_{N}+b_{+}\mathcal{E}c_{N}+f_{+}\mathcal{E}^{\sharp}c_{N}^{\dagger}\right].\label{b6}
		\end {aligned}
	\end{equation}
	The Hermiticity of Hamiltonian (\ref{b6}) requires $b_\pm=f_\pm^{*}$ and $a_\pm \in \mathbb{R}$.
	
	\subsection{Fusion procedure}
	The fusion approach introduced in Section \ref{section2} is also applicable to open systems. In Section \ref{Sec:FusionR}, we have demonstrated the fusion of the $R$-matrices and subsequently applied it to construct the fused monodromy matrices given by Eq. \eqref{FM1}.

    The fused analogues for the reflection monodromy matrix $\hat{T}(u)$ are constructed in the same way, specifically
	\begin{equation}\label{oo2}
	\hat{T}_{\alpha}(u) = R_{N,\alpha}(u+\theta_{N}) \cdots R_{2,\alpha}(u+\theta_{2}) R_{1,\alpha}(u+\theta_{1}), \quad
	\alpha \in \{\bar{0}, \bar{0}^\prime, \tilde{0}, \tilde{0}^\prime\}.
	\end{equation}
	
	%Next, we introduce the fusion of the $K$-matrices and thereby construct the fused transfer matrices for the open system.
	
	\subsubsection{\texorpdfstring{Fused $K$-matrices}{Fused K-matrices}}

	For open systems, we should also perform the fusion procedure of the $K$-matrices using the same projectors as those used for the $R$-matrices, which are introduced in Section \ref{Sec:FusionR}.
	
	The first-level fused $K$-matrices are 
	%In order to obtain the sufficient constraints to determine the eigenvalues of Hamiltonian (\ref{b6}), we consider  the fusion of boundary reflection matrices \cite{ML1992,ZY1996}. The fusion processes should be consistent with that for the $R$-matrices. The 2-dimensional fusions by the super projectors $P_{1,2}^{(+)}$ (\ref{e}) and ${P}_{1,2}^{(-)}$ (\ref{ju1}) give
	\begin {align}\label{f1}
	\begin {aligned}
	K_{\bar{1}}^{-}(u)&=\left[[1+(u-\tfrac{1}{2}\eta)a_{-}](u+\tfrac{1}{2}\eta)\right]^{-1}P_{2,1}^{(+)}K_{1}^{-}(u-\tfrac{1}{2}\eta)R_{2,1}(2u)K_{2}^{-}(u+\tfrac{1}{2}\eta)P_{1,2}^{(+)},\\[3pt]
	K_{\bar{1}}^{+}(u)&=\left[[1+(u+\tfrac{1}{2}\eta)a_{+}](u-\tfrac{1}{2}\eta)\right]^{-1}P_{1,2}^{(+)}K_{2}^{+}(u+\tfrac{1}{2}\eta)R_{1,2}(-2u)K_{1}^{+}(u-\tfrac{1}{2}\eta)P_{2,1}^{(+)},\\[3pt]
	K_{\bar{1}^{\prime}}^{-}(u)&=\left[[1-(u+\tfrac{1}{2}\eta)a_{-}](u-\tfrac{1}{2}\eta)\right]^{-1}{P}_{2,1}^{(-)}K_{1}^{-}(u+\tfrac{1}{2}\eta)R_{2,1}(2u)K_{2}^{-}(u-\tfrac{1}{2}\eta){P}_{1,2}^{(-)},\\[3pt]
	K_{\bar{1}^{\prime}}^{+}(u)&=\left[[1-(u-\tfrac{1}{2}\eta)a_{+}](u+\tfrac{1}{2}\eta)\right]^{-1}{P}_{1,2}^{(-)}K_{2}^{+}(u-\tfrac{1}{2}\eta)R_{1,2}(-2u)K_{1}^{+}(u+\tfrac{1}{2}\eta){P}_{2,1}^{(-)}.
	\end {aligned}
	\end {align}
	The second-level fused $K$-matrices read
	\begin {equation}\label{f2}
	\begin {aligned}
	K_{\tilde{1}}^{-}(u)&=\left[2[1-(u+\eta)a_{-}](u-\tfrac{1}{2}\eta)\right]^{-1}\mathbb{P}_{\bar{1},2}^{(-)}K_{2}^{-}(u+\eta)R_{\bar{1},2}(2u+\tfrac{1}{2}\eta)K_{\bar{1}}^{-}(u-\tfrac{1}{2}\eta)\mathbb{P}_{2,\bar{1}}^{(-)},\\[3pt]
	K_{\tilde{1}}^{+}(u)&=\left[2(1-ua_+)(u+\eta)\right]^{-1}\mathbb{P}_{2,\bar{1}}^{(-)}K_{\bar{1}}^+(u-\tfrac{1}{2}\eta)R_{2,\bar{1}}(-2u-\tfrac{1}{2}\eta)K_2^+(u+\eta)\mathbb{P}_{\bar{1},2}^{(-)},\\[3pt]
	K_{\tilde{1}^{\prime}}^{-}(u)&=\left[2[1+(u-\eta)a_-](u+\tfrac{1}{2}\eta)\right]^{-1}\mathcal{P}_{\bar{1}^{\prime},2}^{(+)}K_{2}^{-}(u-\eta)R_{\bar{1}^{\prime},2}(2u-\tfrac{1}{2}\eta)K_{\bar{1}^{\prime}}^-(u+\tfrac{1}{2}\eta)\mathcal{P}_{2,\bar{1}^{\prime}}^{(+)},\\[3pt]
	K_{\tilde{1}^{\prime}}^{+}(u)&=\left[2(1+ua_{+})(u-\eta)\right]^{-1}\mathcal{P}_{2,\bar{1}^{\prime}}^{(+)}K_{\bar{1}^{\prime}}^{+}(u+\tfrac{1}{2}\eta)R_{2,\bar{1}^{\prime}}(-2u+\tfrac{1}{2}\eta)K_{2}^{+}(u-\eta)\mathcal{P}_{\bar{1}^{\prime},2}^{(+)}.
	\end {aligned}
	\end {equation}
	
	It should be remarked that all fused reflection matrices defined in Eqs. (\ref{f1}) and (\ref{f2}) are $2 \times 2$ matrices in their respective fused spaces, and their matrix elements are operator polynomials in $u$ of degree at most one. The fused $K$-matrices satisfy the following fused (dual) reflection equations 
	\begin{align}
	 R _{\alpha,\beta}(u-v){K^{-}_{\alpha}}(u)R _{\beta,\alpha}(u+v) {K^{-}_{\beta}}(v)&={K^{-}_{\beta}}(v)R _{\alpha,\beta}(u+v){K^{-}_{\alpha}}(u)R _{\beta,\alpha}(u-v),\label{fused:RE}\\
	R_{\alpha,\beta}(v-u)K_\alpha^+(u)R_{\beta,\alpha}(-u-v)K_\beta^+(v)&=K_\beta^+(v)R_{\alpha,\beta}(-u-v)K_\alpha^+(u)R_{\beta,\alpha}(v-u),\label{fused:DRE}
	\end{align}
	where indices $\alpha, \beta$ may label either the original spaces or the projected spaces.
	
	Using Eq.  (\ref{m}), we can finally get 
	\begin {equation}
	K_{\tilde{1}}^{-}(u)=K_{\tilde{1}^{\prime}}^{-}(u), \quad  K_{\tilde{1}}^{+}(u)=K_{\tilde{1}^{\prime}}^{+}(u). \label{f3}
	\end {equation}
	The situation now is quite similar to the fusion of $R$-matrices described in Section \ref{Sec:FusionR}. Specifically, the $K$-matrix fusion also follows two  branches that subsequently interconnect after two fusion levels, as illustrated  in Fig. \ref{fig1} (with $R(u)$ replaced by $K^\pm(u)$).
	
	\subsubsection{Fused transfer matrices}
		The fused transfer matrices are defined as
	\begin{equation}\label{tt1}
	\begin{aligned}
	&t^{(1)}(u)={\rm str}_{\bar{0}}\{K_{\bar{0}}^{+}(u)T_{\bar{0}}(u)K_{\bar{0}}^{-}(u)\hat{T}_{\bar{0}}(u)\},\\
	 &t^{(2)}(u)={\rm str}_{\bar{0}^{\prime}}\{K_{\bar{0}^{\prime}}^{+}(u)T_{\bar{0}^{\prime}}(u)K_{\bar{0}^{\prime}}^{-}(u)\hat{T}_{\bar{0}^{\prime}}(u)\},\\
	&\tilde{t}^{(1)}(u)={\rm str}_{\tilde{0}}\{K_{\tilde{0}}^{+}(u)T_{\tilde{0}}(u)K_{\tilde{0}}^{-}(u)\hat{T}_{\tilde{0}}(u)\},\\
	&\tilde{t}^{(2)}(u)={\rm str}_{\tilde{0}^{\prime}}\{K_{\tilde{0}^{\prime}}^{+}(u)T_{\tilde{0}^{\prime}}(u)K_{\tilde{0}^{\prime}}^{-}(u)\hat{T}_{\tilde{0}^{\prime}}(u)\}.
    \end{aligned}
	\end{equation}
	From Eqs. (\ref{m}), (\ref{f3}), and (\ref{tt1}), it follows that the fused transfer matrices $\tilde{t}^{(1)}(u)$ and $\tilde{t}^{(2)}(u)$ are identical. We therefore denote them collectively as $\tilde{t}(u)$
	\begin{equation}\label{jie2}
	\tilde{t}(u)=\tilde{t}^{(1)}(u)=\tilde{t}^{(2)}(u).
	\end{equation}
	Equations \eqref{a1a}, \eqref{fused:RE} and \eqref{fused:DRE} allow us to prove that $t(u)$, $t^{(1)}(u)$, $t^{(2)}(u)$, and $\tilde{t}(u)$ are mutually commutative.

	\subsection{Operator identities}
	\paragraph{Operator product identities}
	
We introduce the function
\begin{align}
\alpha(u)=(1+ua_{-})[1+(u+\eta)a_{+}]\prod_{j=1}^{N}(u+\theta_{j}+\eta)(u-\theta_{j}+\eta).
\end{align}
The fused transfer matrices defined in Eq. \eqref{tt1} satisfy the following operator product identities
	\begin{equation}
		\begin{aligned}
		&t(\pm\theta_j)t(\pm\theta_j+\eta)=-\frac{1}{4}\frac{\pm\theta_j(\pm\theta_j+\eta)}{(\pm\theta_j+\frac{1}{2}\eta)^2} \, \alpha(\pm\theta_{j})t^{(1)}(\pm\theta_j+\tfrac{1}{2}\eta), \\[5pt]
		&t(\pm\theta_j-\eta)t(\pm\theta_j)=-\frac{1}{4}\frac{\pm\theta_j(\pm\theta_j-\eta)}{(\pm\theta_j-\frac{1}{2}\eta)^2}\, \alpha(\mp\theta_{j})t^{(2)}(\pm\theta_j-\tfrac{1}{2}\eta), \\[5pt]
		&t^{(1)}(\pm\theta_{j}-\tfrac{3}{2}\eta)t(\pm\theta_{j})=-\frac{\pm\theta_{j}(\pm\theta_{j}-\frac{3}{2}\eta)}{(\pm\theta_{j}-\frac{1}{2}\eta)(\pm\theta_{j}-\eta)}
		\, \alpha(\mp\theta_{j})\tilde{t}(\pm\theta_{j}-\eta), \\[5pt]
		&t^{(2)}(\pm\theta_{j}+\tfrac{3}{2}\eta)t(\pm\theta_{j})=-\frac{\pm\theta_{j}(\pm\theta_{j}+\frac{3}{2}\eta)}{(\pm\theta_{j}+\frac{1}{2}\eta)(\pm\theta_{j}+\eta)}
		\, \alpha(\pm\theta_{j})\tilde{t}(\pm\theta_{j}+\eta),
		\end{aligned}\label{o1}
	\end{equation}
	where $j=1,\ldots,N$.
	A detailed proof of (\ref{o1}) is provided in Appendix \ref{APPB}.
	
	\paragraph{Transfer matrices at specific points}
	 The properties of the $R$-matrices and $K$-matrices enable the direct evaluation of transfer matrices at specific points
		\begin{equation}\label{o5}
			\begin{aligned}
				&t(0)=0,\quad{t}^{(1)}(0)=0,\quad{t}^{(2)}(0)=0,\quad\tilde{t}(0)=0, 
				\quad t^{(1)}(-\tfrac{1}{2}\eta)=-2t(-\eta),\\
				&{t}^{(1)}(\tfrac{1}{2}\eta)=-2t(\eta),\quad{t}^{(2)}(-\tfrac{1}{2}\eta)=2t(-\eta), \quad
				t^{(2)}(\tfrac{1}{2}\eta)=2t(\eta),\quad\tilde{t}(\eta)=\tfrac{2}{3}t^{(1)}(\tfrac{3}{2}\eta).
			\end{aligned}
		\end{equation}
		
		\paragraph{Asymptotic behavior}
		Through a straightforward analysis, we obtain the following asymptotic forms of the transfer matrices $t(u)$, $ t^{(1)}(u)$, $ t^{(2)}(u)$ and $\tilde{t}(u)$
		\begin{equation}\label{o6}
			\begin{aligned}
				&t(u)|_{u\to\infty}=2\kappa\, u^{2N+1}\times \mathbb{I}+\cdots,\\	&t^{(1)}(u)|_{u\to\infty}=-8\kappa\, u^{2N+1}\times\mathbb{I}+\cdots,\\
				&t^{(2)}(u)|_{u\to\infty}=8\kappa\, u^{2N+1}\times \mathbb{I}+\cdots,\\
				&\tilde{t}(u)|_{u\to\infty}=-8\kappa\, u^{2N+1}\times \mathbb{I}+\cdots,
			\end{aligned}
		\end{equation}
		where $\kappa=a_{+}+a_{-}+a_{+}a_{-}\eta$.

	\subsection{\texorpdfstring{$T$-$Q$ relation}{TQ relation}}
	
	The transfer matrices $t(u)$, $t^{(1)}(u)$, $t^{(2)}(u)$, and $\tilde{t}(u)$ commute with each other and consequently possess common eigenstates. Let $\Lambda(u)$, $\Lambda^{(1)}(u)$, $\Lambda^{(2)}(u)$, and $\tilde{\Lambda}(u)$ denote their respective eigenvalues. Then, Eqs. \eqref{o1}--\eqref{o6} directly imply
	\begin{equation}
		\begin{aligned}
		&\Lambda(\pm\theta_j)\Lambda(\pm\theta_j+\eta)=-\frac{1}{4}\frac{\pm\theta_j(\pm\theta_j+\eta)}{(\pm\theta_j+\frac{1}{2}\eta)^2} \,
		\alpha(\pm\theta_{j})\Lambda^{(1)}(\pm\theta_j+\tfrac{1}{2}\eta), \\[5pt]
		&\Lambda(\pm\theta_j-\eta)\Lambda(\pm\theta_j)=-\frac{1}{4}\frac{\pm\theta_j(\pm\theta_j-\eta)}{(\pm\theta_j-\frac{1}{2}\eta)^2}
		\, \alpha(\mp\theta_{j})\Lambda^{(2)}(\pm\theta_j-\tfrac{1}{2}\eta),\\[5pt]
		&\Lambda^{(1)}(\pm\theta_{j}-\tfrac{3}{2}\eta)\Lambda(\pm\theta_{j})=-\frac{\pm\theta_{j}(\pm\theta_{j}-\frac{3}{2}\eta)}{(\pm\theta_{j}-\frac{1}{2}\eta)(\pm\theta_{j}-\eta)}
		\, \alpha(\mp\theta_{j})\tilde{\Lambda}(\pm\theta_{j}-\eta),\\[5pt]
		&\Lambda^{(2)}(\pm\theta_{j}+\tfrac{3}{2}\eta)\Lambda(\pm\theta_{j})=-\frac{\pm\theta_{j}(\pm\theta_{j}+\frac{3}{2}\eta)}{(\pm\theta_{j}+\frac{1}{2}\eta)(\pm\theta_{j}+\eta)}\,
		\alpha(\pm\theta_{j})\tilde{\Lambda}(\pm\theta_{j}+\eta),
	\end{aligned}\label{fun1}
	\end{equation}
	where $j=1,2,\ldots,N$ and
	%{\color{blue} According to Eqs. (\ref{o5}) and (\ref{o6}) , we also have another 13 constraints as}
	\begin{align}\label{o8}
		&\begin{aligned}
			&\Lambda(0)=0,\quad{\Lambda}^{(1)}(0)=0,\quad{\Lambda}^{(2)}(0)=0,\quad\tilde{\Lambda}(0)=0,\\
			&\Lambda^{(1)}(-\tfrac{1}{2}\eta)=-2\Lambda(-\eta),\qquad {\Lambda}^{(1)}(\tfrac{1}{2}\eta)=-2\Lambda(\eta), \\
			&{\Lambda}^{(2)}(-\tfrac{1}{2}\eta)=2\Lambda(-\eta),\quad \Lambda^{(2)}(\tfrac{1}{2}\eta)=2\Lambda(\eta),\quad\tilde{\Lambda}(\eta)=\tfrac{2}{3}\Lambda^{(1)}(\tfrac{3}{2}\eta),
			\end{aligned}
			\\[2pt]
			&\begin{aligned}\label{o9}
			&\Lambda(u)|_{u\to\infty}=2\kappa \, u^{2N+1} +\cdots,\qquad \Lambda^{(1)}(u)|_{u\to\infty}=-8\kappa\,u^{2N+1} +\cdots ,\\
			&\Lambda^{(2)}(u)|_{u\to\infty}=8\kappa \, u^{2N+1} +\cdots,\qquad \tilde{\Lambda}(u)|_{u\to\infty}=-8\kappa \,u^{2N+1} +\cdots.
		\end{aligned}
	\end{align}
	
	From the definitions of the transfer matrices in Eqs. \eqref{b5} and \eqref{tt1}, we know that $\Lambda(u)$, $\Lambda^{(1)}(u)$, $\Lambda^{(2)}(u)$, and $\tilde{\Lambda}(u)$ are all polynomials in $u$ of degree $2N+2$. 
    The $8N+13$ equations in \eqref{fun1} - \eqref{o9} thus provide sufficient constraints to determine these functions completely.
	
	We can parameterize $\Lambda(u)$, $\Lambda^{(1)}(u)$, $\Lambda^{(2)}(u)$, and $\tilde{\Lambda}(u)$ by the following $T$-$Q$ relations
	\begin{equation}\label{TQ:open}
		\begin{aligned}
			&\Lambda(u)=\frac{2u}{2u+\eta}\left[\alpha(u)-\alpha(-u-\eta)\right]\frac{Q(u-\eta)}{Q(u)},\\
			&\Lambda^{(1)}(u)=-\frac{4u}{u+\eta}\left[\alpha(u+\tfrac{\eta}{2})-\alpha(-u-\tfrac{3}{2}\eta)\right]\frac{Q(u-\tfrac{3\eta}{2})}{Q(u+\tfrac{\eta}{2})},\\
			&\Lambda^{(2)}(u)=\frac{4u}{u+\eta}\left[\alpha(u+\tfrac{\eta}{2})-\alpha(-u-\tfrac{3}{2}\eta)\right]\frac{Q(u-\tfrac{3\eta}{2})}{Q(u+\tfrac{\eta}{2})},\\
			&\tilde{\Lambda}(u)=-\frac{8u}{2u+3\eta}\left[\alpha(u+\eta)-\alpha(-u-2\eta)\right]\frac{Q(u-2\eta)}{Q(u+\eta)},
		\end{aligned}
	\end{equation}
	where
	\begin{alignat}{2}
	Q(u)=\prod_{k=1}^{M}(u-\lambda_{k})(u+\lambda_{k}+\eta),\quad 0\leq M\leq N.\label{def:Q:open}
	\end{alignat}
	The Bethe roots $\{\lambda_1,\ldots,\lambda_M\}$ satisfy the following BAEs
	\begin{equation}
		\frac{\alpha(\lambda_{k})}{\alpha(-\lambda_{k}-\eta)}=1,\quad k=1,\ldots M.\label{BAE:open}
	\end{equation}

	The eigenvalue of the Hamiltonian (\ref{b6}) in terms of the Bethe roots  is given by 
	\begin{equation}\label{en2}
		\begin{aligned}
			E&=\frac{1}{8\eta^{N}(1+a_{+}\eta)} \left.\frac{\partial ^2 \Lambda(u)}{\partial u^2}\right|_{u=0,\{\theta_j=0\}}\\
			&=\eta^{N}\sum_{k=1}^{M}\frac{1}{\lambda_{k}(\lambda_{k}+\eta)}+\frac{\eta^{N-2}}{2}\left(2N-1+a_{-}\eta-\frac{1}{1+a_{+}\eta}\right).
		\end{aligned}
	\end{equation}
	
	Numerical results for the Bethe roots with system size $N=3$ are presented in Table \ref{tab3}. 
	We note that the eigenvalue of the Hamiltonian derived from the Bethe roots coincides with that given by the direct diagonalization of the Hamiltonian.
	
	\begin{table}[htbp]
		\caption{Numeric results of Bethe roots $\{\lambda_k\}$ and eigenvalues of the Hamiltonian (\ref{b6}) with $N=3$, $\eta=1$ and $a_+=0.5,a_-=1.2$ and $\{\theta_j=0\}$.}\label{tab3}
		\centering
		\vspace{\baselineskip}
		\renewcommand{\arraystretch}{1.5}
		\begin{tabular}{|c|c|c|c|}
			\hline
			$\lambda_1$ & $\lambda_2$ &  $\lambda_3$& $E$ \\
			\hline
			--    &  --  & -- &   \makebox[3.5em][r]{2.7667}\\
			$-$0.5000$-$1.5235i     & --  & -- &  \makebox[3.5em][r]{2.3777}\\
			$-$0.5000$-$0.2187i  & --  & -- &  $-$0.5911\\
			$-$0.5000$-$0.5565i  & --  & -- &  \makebox[3.5em][r]{0.9800}\\
			$-$0.5000$-$1.5235i &$-$0.5000$-$0.2187i  &   -- &  $-$0.9800\\
			$-$0.5000$-$1.5235i &$-$0.5000$-$0.5565i  & -- &  \makebox[3.5em][r]{0.5911}\\
			$-$0.5000$-$0.2187i &$-$0.5000$-$0.5565i & -- &  $-$2.3777\\
			$-$0.5000$-$1.5235i &$-$0.5000$-$0.2187i&$-$0.5000$-$0.5565i &  $-$2.7667\\
			\hline
		\end{tabular}
	\end{table}
	Since Grassmann numbers are absent from equations \eqref{fun1} - \eqref{o9}, it follows directly that the eigenvalues of the transfer matrix and the Hamiltonian are independent of them. In contrast, the eigenstates are strongly dependent on these Grassmann numbers.
	
	We observe that the presence of boundary Grassmann numbers breaks the $U(1)$ symmetry of the system. Nevertheless, the $T$-$Q$ relations in Eq. \eqref{TQ:open} share similar structures to the ones in the periodic case (Eq. \eqref{TQ:periodic}). The $T$-$Q$ relation in Eq. \eqref{TQ:open} matches the earlier conjecture in Ref. \cite{GAM2010}, which was only checked numerically for small systems without an analytic proof. We address this problem by obtaining the relation analytically via the fusion approach.
	
	The derivation of the exact spectrum of the model allows us to retrieve the Bethe state, which we will demonstrate in the following section.

\section{\texorpdfstring{Bethe state of the open $\mathfrak{gl}(1|1)$ integrable model}{Bethe state of the open gl(1|1) integrable model}}	
	\label{section4}
The Bethe-type  eigenstates of integrable models with generic open boundary conditions can be constructed \cite{ODBA_book,zhang15,Zhang2014,Belliard2014,Belliard:2014,Wen2021}. In this work, we apply the approach in Refs. \cite{zhang15,Zhang2014} to retrieve the Bethe states of the open $\mathfrak{gl}(1|1)$ model. By employing two sets of gauge transformations, we obtain appropriate generators and a reference state for constructing the Bethe vectors, respectively. To verify the Bethe state, we also construct a complete basis of the Hilbert space via the separation of variables (SoV) approach \cite{Niccoli2012_1,Niccoli2012_2,Kitanine2015}.

\subsection{Gauge transformation}	
For convenience, we denote the double-row monodromy matrix as
\begin{eqnarray}
\mathscr{U}(u)=T(u)K^{-}(u)\hat{T}(u)=\begin{pmatrix}
\mathscr{A}(u)&\mathscr{B}(u)\\
\mathscr{C}(u)&\mathscr{D}(u)\end{pmatrix}.
\end{eqnarray}
The transfer matrix $t(u)$ in Eq. (\ref{b5}) can be expressed as a linear combination of the elements of  double-row monodromy matrix
\begin{align}
t(u)&=K^{+}_{11}(u)\mathscr{A}(u)+K^{+}_{12}(u)\mathscr{C}(u)-K^{+}_{21}(u)\mathscr{B}(u)-K^{+}_{22}(u)\mathscr{D}(u).
\end{align}

The reflection matrix $K^{+}(u)$ (\ref{b3}) can be diagonalized as follows 
\begin{eqnarray}
	&&\tilde{K}^{+}(u)=\tilde GK^{+}(u)\tilde G^{-1}=\begin{pmatrix}\tilde{K}^{+}_{11}(u)&0\\0&\tilde{K}^{+}_{22}(u)\end{pmatrix}=\begin{pmatrix}1+ua_{+}&0\\0&1-ua_{+}\nonumber\\ \end{pmatrix},
\end{eqnarray}
where the gauge transformation matrix $\tilde G$ and its reverse $\tilde G^{-1}$ are
\begin{eqnarray}
\tilde G=\frac{1}{2a_{+}}\begin{pmatrix}
	2a_{+}&b_{+}\boldsymbol{\mathcal{E}} 
\\-f_{+}\boldsymbol{\mathcal{E}}^{\sharp}&2a_{+}\end{pmatrix},\qquad 	\tilde G^{-1}=\frac{1}{2a_{+}}\begin{pmatrix}
2a_{+}&-b_{+}\boldsymbol{\mathcal{E}} 
\\f_{+}\boldsymbol{\mathcal{E}}^{\sharp}&2a_{+}\end{pmatrix}.\label{g1}
\end{eqnarray}
By applying the same gauge transformation to the $R$-matrices and $K^-(u)$, we arrive at
\begin{align}
t(u)={\rm str}_0\{\tilde K_0^+(u)\tilde {\mathscr U}(u)\}=\tilde{K}^{+}_{11}(u)\tilde{\mathscr{A}}(u)-\tilde{K}^{+}_{22}(u)\tilde{\mathscr{D}}(u),
\end{align}
where 
\begin{align}
\tilde{\mathscr{U}}(u)=\tilde{G}T(u)K^{-}(u)\hat{T}(u)\tilde{G}^{-1}=\tilde GT(u)\tilde G^{-1}\tilde K^{-}(u) \tilde{G}\hat{T}(u)\tilde G^{-1}=\begin{pmatrix}
\tilde{\mathscr{A}}(u)&\tilde{\mathscr{B}}(u)\\
\tilde{\mathscr{C}}(u)&\tilde{\mathscr{D}}(u)\end{pmatrix},\label{gauge:U}
\end{align}
and $\tilde{K}^{-}(u)$ is defined as \begin{align}
\tilde{K}^{-}(u)&=\tilde GK^{-}(u)\tilde G^{-1}= \begin{pmatrix}\tilde{K}^{-}_{11}(u)&\tilde{K}^{-}_{12}(u)\\\tilde{K}^{-}_{21}(u)&\tilde{K}^{-}_{22}(u)\end{pmatrix}\nonumber\\
&=\frac{1}{a_{+}}\begin{pmatrix}a_{+}(1+ua_{-})&(a_{+}b_{-}-b_{+}a_{-})u\boldsymbol{\mathcal{E}}\\(a_{+}f_{-}-a_{-}f_{+})u\boldsymbol{\mathcal{E}}^{\sharp}&a_{+}(1-ua_{-})\end{pmatrix}.\label{gauge:K-}
\end{align}
The entries of $\mathscr{U}(u)$ and $\tilde{\mathscr{U}}(u)$ satisfy the following relations 
\begin{align}
\begin{aligned}
&\tilde{\mathscr{A}}(u)=\mathscr{A}(u)-\frac{f_{+}}{2a_{+}}\boldsymbol{\mathcal{E}}^{\sharp}\mathscr{B}(u)+\frac{b_{+}}{2a_{+}}\boldsymbol{\mathcal{E}}\mathscr{C}(u),\\
&\tilde{\mathscr{B}}(u)=-\frac{b_{+}}{2a_{+}}\boldsymbol{\mathcal{E}}\Big[\mathscr{A}(u)-\mathscr{D}(u)\Big]+\mathscr{B}(u),\\
&\tilde{\mathscr{C}}(u)=-\frac{f_{+}}{2a_{+}}\boldsymbol{\mathcal{E}}^{\sharp}\Big[\mathscr{A}(u)-\mathscr{D}(u)\Big]+\mathscr{C}(u),\\
&\tilde{\mathscr{D}}(u)=-\frac{f_{+}}{2a_{+}}\boldsymbol{\mathcal{E}}^{\sharp}\mathscr{B}(u)+\frac{b_{+}}{2a_{+}}\boldsymbol{\mathcal{E}}\mathscr{C}(u)+\mathscr{D}(u).\label{eq:elements}
\end{aligned}
\end{align}

It is important to note that Grassmann numbers commute with the diagonal elements, but anti-commute with the off-diagonal elements, of the double-row monodromy matrix—a property that holds for both its original and gauge-transformed versions.

The commutation relations among $\tilde{\mathscr{A}}(u), \tilde{\mathscr{B}}(u), \tilde{\mathscr{C}}(u), \tilde{\mathscr{D}}(u)$ are the same as those among the untransformed operators. A number of useful specific relations are provided in Appendix \ref{app:D}.

%\begin{align}
%&\bar{\mathscr{A}}(u)=\mathscr{A}(u)-\frac{f_{-}}{2a_{-}}\boldsymbol{\mathcal{E}}^{\sharp}\mathscr{B}(u)+\frac{b_{-}}{2a_{-}}\boldsymbol{\mathcal{E}}\mathscr{C}(u),\nonumber\\
%&\bar{\mathscr{B}}(u)=-\frac{b_{-}}{2a_{-}}\boldsymbol{\mathcal{E}}\Big[\mathscr{A}(u)-\mathscr{D}(u)\Big]+\mathscr{B}(u),\nonumber\\
%&\bar{\mathscr{C}}(u)=-\frac{f_{-}}{2a_{-}}\boldsymbol{\mathcal{E}}^{\sharp}\Big[\mathscr{A}(u)-\mathscr{D}(u)\Big]+\mathscr{C}(u),\nonumber\\
%&\bar{\mathscr{D}}(u)=-\frac{f_{-}}{2a_{-}}\boldsymbol{\mathcal{E}}^{\sharp}\mathscr{B}(u)+\frac{b_{-}}{2a_{-}}\boldsymbol{\mathcal{E}}\mathscr{C}(u)+\mathscr{D}(u).
%\end{align}

\subsection{SoV Basis}
To begin, we rewrite the $R$-matrices $R_{0,j}(u)$, $R_{j,0}(u)$  in the auxiliary space $V_0$ 
\begin{eqnarray}
R_{0,j}(u)=R_{j,0}(u)=\begin{pmatrix}
		u+\eta\,\bar{n}_j&\eta\, c^{\dagger}_j\\\eta \,c_j&u-\eta\, n_j\end{pmatrix},
\end{eqnarray}
where ${c_j,\, c_j^\dagger}$ and $n_k$ denote the fermionic annihilation, creation, and particle number operators, respectively.
By applying the gauge transformation $\tilde G$ to the Lax operator, we obtain
\begin{eqnarray}
\tilde{R}_{0,j}(u)=\tilde G_0R_{0,j}(u)\tilde G_0^{-1}=\begin{pmatrix}
u+\eta\,\tilde{\bar{n}}_j&\eta\,\tilde{c}^{\dagger}_j
\\\eta\,\tilde{c}_j&u-\eta\,\tilde{n}_j\end{pmatrix},
\end{eqnarray}
where 
\begin{eqnarray}
\tilde{\bar{n}}_j=1-n_j
+\rho_1c_j-\rho_2c^{\dagger}_j,\quad \tilde{c}_j^{\dagger}=c_j^{\dagger}-\rho_1,\quad \tilde{c}_j=c_j-\rho_2,\quad \tilde{n}_j=n_j-\rho_1c_j+\rho_2c^{\dagger}_j, 
\end{eqnarray}
and 
\begin{eqnarray}
	\rho_1=\frac{b_{+}\boldsymbol{\mathcal{E}}}{2a_{+}},\qquad \rho_2=\frac{f_{+}\boldsymbol{\mathcal{E}}^{\sharp}}{2a_{+}}.
\end{eqnarray}

Let us introduce the following local state on site $n$
\begin{eqnarray}
|\tilde 0\rangle_n=|0\rangle_n-\rho_2|1\rangle_n,\quad n=1,\cdots, N,
\end{eqnarray}
which satisfies 
\begin{align}
\left[\tilde{R}_{0,j}(u)\right]_{2,1}|\tilde 0\rangle_n=0,\quad \left[\tilde{R}_{0,j}(u)\right]_{1,1}|\tilde 0\rangle_n=(u+\eta)|\tilde 0\rangle_n,\quad \left[\tilde{R}_{0,j}(u)\right]_{2,2}|\tilde 0\rangle_n=u|\tilde 0\rangle_n.
\end{align}
Analogously, the following local bra vector can also be constructed
\begin{eqnarray}
	_{n}\langle\tilde{0}|= {_{n}\langle}0|-{_{n}\langle1|}\rho_{1},\quad  n=1,\cdots, N,
\end{eqnarray}
which satisfies 
\begin{align}
{_n\langle}\tilde 0|\left[\tilde{R}_{0,j}(u)\right]_{1,2}=0,\quad {_n\langle}\tilde 0|\left[\tilde{R}_{0,j}(u)\right]_{1,1}={_n\langle}\tilde 0|(u+\eta),\quad {_n\langle}\tilde 0|\left[\tilde{R}_{0,j}(u)\right]_{2,2}={_n\langle}\tilde 0|u.
\end{align}
We then introduce two global product states
\begin{eqnarray}
|\omega_0\rangle=|\tilde 0\rangle_{1}\otimes_s|\tilde 0\rangle_{2}\cdots\otimes_{s}|\tilde 0\rangle_{N},\qquad \langle{\omega}_0|={_{1}\langle}\tilde 0|\otimes_s {_{2}\langle}\tilde 0|\cdots \otimes_s{_{N}\!\langle}\tilde 0|.
\end{eqnarray}
From the definition of the gauged double-row monodromy matrix, it can be shown that $|\omega_0\rangle$ and $\langle \omega_0|$ are eigenstates of $\tilde{\mathscr{C}}(u)$ and $\tilde{\mathscr{B}}(u)$, respectively
\begin{eqnarray}
&&\tilde{\mathscr{C}}(u)|\omega_0\rangle=\tilde{K}_{21}^{-}(u)w_-(u)w_+(u+\eta)|\omega_0\rangle,\\
&&\langle{\omega_0}|\tilde{\mathscr{B}}(u)=\langle{\omega_1}|\tilde{K}_{12}^{-}(u)w_-(u+\eta)w_+(u),
\end{eqnarray}
where 
\begin{align}
w_{\pm}(u)=\prod_{j=1}^N(u\pm\theta_j).
\end{align}
Let's construct the SoV vectors
\begin{eqnarray}
&&|{p_1},\dots,{p_n}\rangle=\tilde{\mathscr{A}}(\theta_{p_1})\dots\tilde{\mathscr{A}}(\theta_{p_n})|\omega_0\rangle,\label{sov:1}
%&&\langle{q_1},\dots,{q_n}|=\langle\bar{\Omega}|\tilde{\mathscr{D}}(-\theta_{q_1})\dots\tilde{\mathscr{D}}(-\theta_{q_n}),\label{sov:2}
\end{eqnarray}
where $p_j\in\{1,\dots,N\},\, p_1<p_2<\cdots<p_n$. %and $q_1<q_2<\cdots<q_n$. 
With the help of the following identity  
\begin{eqnarray}
&&\tilde{\mathscr{C}}(\theta_j)|\omega_0\rangle=0,\label{com-3}%=\tilde{\mathscr{C}}(-\theta_j-\eta)|\omega_0\rangle=0,
%	&&\langle\bar{\Omega}|\tilde{\mathscr{C}}(-\theta_j)=\langle\bar{\Omega}|\tilde{\mathscr{C}}(\theta_j-\eta)=0,\label{com-4}
\end{eqnarray}
and Eqs. (\ref{com-1}), (\ref{com-2}), we can prove that the vectors defined in Eqs. (\ref{sov:1}) %(\ref{sov:2}) 
are all the eigenstates of $\tilde{\mathscr{C}}(u)$
\begin{eqnarray}
	&&\tilde{\mathscr{C}}(u)|{p_1},\dots,{p_n}\rangle=h(u,\{{p_1},\dots,{p_n}\})|{p_1},\dots,{p_n}\rangle,\label{sov-1}
%	&&\langle{q_1},\dots,{q_n}|\tilde{\mathscr{C}}(u)=\bar{h}(u,\{{q_1},\dots,{q_n}\})\langle{q_1},\dots,{q_n}|,
\end{eqnarray}
with the corresponding eigenvalues being
\begin{eqnarray}
&&	h(u,\{{p_1},\dots,{p_n}\})=\tilde{K}_{21}^{-}(u)w_-(u)w_+(u+\eta)\prod_{l=1}^{n}\frac{(u+\theta_{p_l})(u-\theta_{p_l}+\eta)}{(u-\theta_{p_l})(u+\theta_{p_l}+\eta)}.\label{sov-2}
%&&	\bar{h}(u,\{{q_1},\dots,{q_n}\})=\tilde{K}_{21}^{-}(u)w_-(u+\eta)w_+(u)\prod_{l=1}^{n}\frac{(u-\theta_{q_l})(u+\theta_{q_l}-\eta)}{(u-\theta_{q_l})(u-\theta_{q_l}-\eta)}.
\end{eqnarray}
We see that the vector $|{p_1},\dots,{p_n}\rangle$ does not depend on the order of $\tilde{\mathscr{A}}(\theta_{p_j})$, i.e., $$|\dots,{p_j},\dots,{p_k},\dots\rangle=|\dots,{p_k},\dots,{p_j},\dots\rangle.$$
Furthermore, vectors $\{|{p_1},\dots,{p_n}\rangle\}$ with distinct configurations $\{p_1,\ldots,p_n\}$ are mutually orthogonal due to the difference in their corresponding spectra. As the total number of the SoV vectors in (\ref{sov-1}) equals the Hilbert space dimension, they form a complete basis.

Similarly, we can construct another set of Sov basis of the Hilbert
\begin{align}
\langle{p_1},\dots,{p_n}|=\langle\omega_0|\tilde{\mathscr{A}}(-\theta_{p_1})\dots\tilde{\mathscr{A}}(-\theta_{p_n}),\label{sov:2}
\end{align}
where $p_j\in\{1,\dots,N\},\, p_1<p_2<\cdots<p_n$. It can be proved that the vectors in Eq. (\ref{sov:2}) all are eigenstates of $\tilde{\mathscr{B}}(u)$.

\subsection{The Scalar Product \texorpdfstring{$\langle\Psi|{p_1},\ldots,{p_n}\rangle$}{<Psi|p1,..pn>}}

We introduce the scalar product 
\begin{eqnarray}
	F_n({p_1},\dots,{p_n})=\langle\Psi|{p_1},\dots,{p_n}\rangle,
\end{eqnarray}
where $\langle\Psi|$ is a common eigenstate of the transfer matrix $t(u)$.
By inserting an operator \(t(\theta_{p_{n+1}})\) between the bra vector \(\langle\Psi|\) and the ket vector \(|p_1,\dots,p_n\rangle\), and alternately acting it to the left and to the right, we obtain the following relation
\begin{align}
&\Lambda(\theta_{p_{n+1}})F_n({p_1},\dots,{p_n})\nonumber\\
&=\tilde{K}_{11}^{+}(\theta_{p_{n+1}})F_{n+1}({p_1},\dots,p_n,{p_{n+1}})-\tilde{K}_{22}^{+}(\theta_{p_{n+1}})\langle\Psi|\tilde{\mathscr{D}}(\theta_{p_{n+1}})\prod_{l=1}^{n}\tilde{\mathscr{A}}(\theta_{p_{l}})|\omega_0\rangle.
\end{align}
Introduce a useful identity 
\begin{align}
\tilde{\mathscr{D}}(\theta_{k})|\omega_0\rangle=\frac{\eta}{2\theta_k+\eta}\tilde{\mathscr{A}}(\theta_{k})|\omega_0\rangle,\quad k=1,\ldots,N,\label{D:omega}
\end{align}
The commutation relations (\ref{com-2}), together with Eqs. (\ref{sov-1}), (\ref{sov-2}) and (\ref{D:omega}) lead to the following identity
\begin{align}
&\tilde{\mathscr{D}}(\theta_{p_{n+1}})\prod_{l=1}^{n}\tilde{\mathscr{A}}(\theta_{p_{l}})|\omega_0\rangle=\prod_{l=1}^{n}\tilde{\mathscr{A}}(\theta_{p_{l}})\tilde{\mathscr{D}}(\theta_{p_{n+1}})|\omega_0\rangle\nonumber\\
&=\frac{\eta}{2\theta_{p_{n+1}}+\eta}\prod_{l=1}^n\tilde{\mathscr{A}}(\theta_{p_{l}})\tilde{\mathscr{A}}(\theta_{p_{n+1}})|\omega_0\rangle=\frac{\eta}{2\theta_{p_{n+1}}+\eta}|p_1,\dots,p_n,p_{n+1}\rangle.\label{eq:DA}
\end{align}
%By using the commutation relations (\ref{com-5}), we can directly derive the equation\begin{eqnarray}\tilde{\mathscr{D}}(\theta_{p_{j}})|\Omega\rangle=\frac{\eta}{2\theta_{p_j}+\eta}\tilde{\mathscr{A}}(\theta_{p_{j}})|\Omega\rangle,\quad j=1,\cdots, N\label{sov-3}.\end{eqnarray}Substituting (\ref{sov-3}) into the relation (\ref{sov-4}), 
Therefore, we obtain
\begin{eqnarray}
	\Lambda(\theta_{p_{n+1}})F_n({p_1},\dots,{p_n})=\frac{(2\theta_{p_{n+1}}+\eta)\tilde{K}_{11}^{+}(\theta_{p_{n+1}})-\eta\tilde{K}_{22}^{+}(\theta_{p_{n+1}})}{2\theta_{p_{n+1}}+\eta}F_{n+1}({p_1},\dots,{p_{n+1}}),
\end{eqnarray}
which allows us to get the expression of   $\{F_n({p_1},\dots,{p_n})\}$
\begin{eqnarray}
F_n({p_1},\dots,{p_n})=\prod_{l=1}^{n}\frac{(2\theta_{p_l}+\eta)\Lambda(\theta_{p_l})}{(2\theta_{p_l}+\eta)\tilde{K}_{11}^{+}(\theta_{p_{l}})-\eta\tilde{K}_{22}^{+}(\theta_{p_l})}F_0,
\end{eqnarray}
where $F_0=\langle\Psi|\omega_0\rangle$ is an overall factor. 
Substituting the explicit expression of the eigenvalue $\Lambda(u)$ given by $T$-$Q$ relation (\ref{TQ:open}), we further derive
\begin{align}
F_n({p_1},\dots,{p_n})
=\prod_{l=1}^{n}(1+\theta_{p_l}a_{-})w_-(\theta_{p_l}+\eta)w_+(\theta_{p_l}+\eta)\frac{Q(\theta_{p_l}-\eta)}{Q(\theta_{p_l})}F_0,\label{Exp:F}
\end{align}
where $Q(u)$ is defined in Eq. (\ref{def:Q:open}). Since the SoV basis is complete, the set $\{F_n({p_1},\dots,{p_n})\}$ can completely determine the form of Bethe state $\langle\Psi|$.
\subsection{Bethe state}
Introduce another gauge transformation 
\begin{align}
\bar{G} = \tilde{G}\big|_{\{a_+,b_+,f_+\} \to \{a_-,b_-,f_-\}},
\end{align}
so that \(K^-(u)\) becomes diagonal under this transformation
\begin{eqnarray}
	&&	\bar{K}^{-}(u)=\bar GK_{-}(u)\bar G^{-1}= \begin{pmatrix}\bar{K}^{-}_{11}(u)&0\\0&\bar{K}^{-}_{22}(u)\end{pmatrix}=\begin{pmatrix}1+ua_{-}& 0\\0&1-ua_{-}\end{pmatrix}.
\end{eqnarray}
Applying the same gauge transformation to the double-row monodromy matrix yields
\begin{align}
\bar{\mathscr{U}}(u) = \bar G \mathscr{U} G^{-1}
= \begin{pmatrix}
\bar{\mathscr{A}}(u) & \bar{\mathscr{B}}(u) \\
\bar{\mathscr{C}}(u) & \bar{\mathscr{D}}(u)
\end{pmatrix}. \label{gauge-1}
\end{align}
Define the following global vectors 
\begin{align}
|\bar\omega_0\rangle=|\omega_0\rangle_{\{a_+,b_+,f_+\} \to \{a_-,b_-,f_-\}},\quad \langle\bar\omega_0|=\langle\omega_0|_{\{a_+,b_+,f_+\} \to \{a_-,b_-,f_-\}}.\label{ref:states}
\end{align}
The state $\langle\bar\omega_0|$ in (\ref{ref:states}) satisfies
\begin{align}
\begin{aligned}
&\langle\bar\omega_0|\bar{\mathscr B}(u)=0,\quad 
\langle\bar\omega_0|\bar{\mathscr A}(u)=\bar{K}_{11}^{-}(u)w_-(u+\eta)w_+(u+\eta)\langle\bar \omega_0|.\label{eq:proof:1}
\end{aligned}
\end{align}
%Next, we want to calculate the overlap between $\langle\bar\omega_0|$ and the SoV basis. 
The aforementioned equation (\ref{eq:proof:1}) together with two other identities
\begin{align}
&\langle\bar\omega_0|\tilde{\mathscr C}(\theta_k)|p_1,\ldots,p_n\rangle=0,\quad k\notin\{p_1,\ldots,p_n\},\\
&\tilde{\mathscr{A}}(u)%=\left(\frac{f_{-}}{2a_{-}}-\frac{f_{+}}{2a_{+}}\right)\boldsymbol{\mathcal{E}}^{\sharp}\mathscr{B}(u)+\left(\frac{b_{+}}{2a_{+}}-\frac{b_{-}}{2a_{-}}\right)\boldsymbol{\mathcal{E}}\mathscr{C}(u)+\bar{\mathscr{A}}(u)\nonumber\\
=\left(\frac{f_{-}}{2a_{-}}-\frac{f_{+}}{2a_{+}}\right)\boldsymbol{\mathcal{E}}^{\sharp}\bar{\mathscr{B}}(u)+\left(\frac{b_{+}}{2a_{+}}-\frac{b_{-}}{2a_{-}}\right)\boldsymbol{\mathcal{E}}\tilde{\mathscr{C}}(u)+\bar{\mathscr{A}}(u),\label{Eq:AA}
\end{align}
allow us to get \begin{align}
 &\langle\bar\omega_0|{p_1},\ldots,{p_n},{p_{n+1}}\rangle= \langle\bar\omega_0|\tilde{\mathscr A}(\theta_{n+1})|{p_1},\ldots,{p_n}\rangle\nonumber\\
 &=\langle\bar \omega_0|\bar{\mathscr A}(\theta_{n+1})|{p_1},\ldots,{p_n}\rangle\nonumber\\
 &=\bar{K}^-_{11}(\theta_{p_{n+1}})w_+(\theta_{p_{n+1}}+\eta)w_-(\theta_{p_{n+1}}+\eta)\langle\bar\omega_0|{p_1},\ldots,{p_n}\rangle.\label{eq:recursive}
\end{align}
Furthermore, we can derive the expression of the overlap $\langle\bar\omega_0|{p_1},\ldots,{p_n}\rangle$ from the recursive relation  (\ref{eq:recursive})
\begin{align}
\langle\bar\omega_0|{p_1},\ldots,{p_n}\rangle=\prod_{k=1}^n\bar{K}^-_{11}(\theta_{p_{k}})w_+(\theta_{p_{k}}+\eta)w_-(\theta_{p_{k}}+\eta)\langle\bar\omega_0|\omega_0\rangle.\label{sov:omega}
\end{align}

\paragraph{Bethe state} The left Bethe state can be parameterized as
\begin{eqnarray}
	\langle\lambda_1,\dots,\lambda_N|=\langle\bar\omega_0|\prod_{l=1}^{M}\tilde{\mathscr{C}}(\lambda_l),\label{BS:left}
\end{eqnarray}
where $\{\lambda_1,\ldots,\lambda_N\}$ are the Bethe roots satisfying BAEs (\ref{BAE:open}), the generator $\tilde{\mathscr{C}}(u)$ and the reference state $\langle\bar\omega_0|$ are defined in Eqs. (\ref{gauge:U}) and (\ref{ref:states}) respectively.

The proof of our Bethe state is straightforward. 
A combination of Eqs. (\ref{sov-1}), (\ref{sov-2}), and (\ref{sov:omega}) yields
\begin{align}
\langle\lambda_1,\dots,\lambda_N|p_1,\ldots,p_n\rangle&=\prod_{l=1}^{n}(1+\theta_{p_l}a_{-})w_-(\theta_{p_l}+\eta)w_+(\theta_{p_l}+\eta)\frac{Q(\theta_{p_l}-\eta)}{Q(\theta_{p_l})}\nonumber\\
&\quad \times \prod_{k=1}^{M}\bar{K}_{21}^{-}(\lambda_{k})w_-(\lambda_k)w_+(\lambda_k+\eta)\langle\bar\omega_0|\omega_0\rangle.\label{sov:Bethe:state}
\end{align}
The factor on the second line of Eq. (\ref{sov:Bethe:state}) is a normalization factor. 
By comparing Eqs. (\ref{Exp:F}) and (\ref{sov:Bethe:state}), we can conclude that $\langle\lambda_1,\dots,\lambda_N|$ is an eigenstate of the transfer matrix.

Analogously, the right Bethe state can also be constructed
\begin{eqnarray}
|\lambda_1,\dots,\lambda_N\rangle=\prod_{l=1}^{M}\tilde{\mathscr{B}}(\lambda_l)|\bar\omega_0\rangle.\label{BS:right}
\end{eqnarray}

It should be remarked that the generation operators, the Bethe roots and the reference states in Eqs. (\ref{BS:left}) and Eqs. (\ref{BS:right}) all have well-defined homogeneous limits of $\{\theta_j\to0\}$. 

Under the condition $a_-f_+ = f_-a_+$, the state $|\bar\omega_0\rangle$ reduces to $|\omega_0\rangle$, and the resulting Bethe state (\ref{BS:right}) coincides with the one given in Ref. \cite{GAM2010}. In this case, we can use the gauge matrix $\tilde G$ to simultaneously diagonalize $K^+(u)$ and triangularize $K^-(u)$ (see Eq. (\ref{gauge:K-})), making the conventional algebraic Bethe ansatz applicable.

	\section{Conclusion}
	\label{section5}
	
	The exact solution of the supersymmetric $\mathfrak{gl}(1|1)$ integrable models with both periodic and generic non-diagonal open boundary conditions is presented in this paper. Using the fusion procedure, we construct a hierarchy of fused transfer matrices, from which a closed set of operator identities is derived. These identities yield the energy spectrum of the model, including the $T$-$Q$ relation and the corresponding Bethe ansatz equations. With the exact spectrum obtained, we then construct the corresponding Bethe states, notably for the open chain with generic non-diagonal boundary conditions.

	The method developed in this work can be applied to other quantum integrable models associated with Lie superalgebra. In particular, it extends straightforwardly to the $U_{q}(\mathfrak{gl}(1|1))$ quantum algebra, for which the $R$–matrix and the reflection $K$–matrices retain the same graded structure as those of the undeformed $\mathfrak{gl}(1|1)$ superalgebra \cite{Zhao06}. In a parallel investigation of the quantum integrable model associated with the Lie superalgebra $\mathfrak{gl}(2|2)$, we have succeeded in establishing virtually all of the operator identities. For higher rank cases, the fusion procedure involves additional levels and branching structures.

	\section*{Acknowledgments}
	
	We thank Prof. Wen-Li Yang for valuable discussions. Financial supports from the National Key R\&D Program of China (Grant No. 2021YFA1402104), National Natural Science Foundation of China (Grant Nos. 12105221, 12247103, 12074410, 12047502, 12434006, 12575007), Shaanxi Fundamental Science Research Project for Mathematics and Physics (Grant Nos. 22JSZ005), Scientific Research Program Funded by Shaanxi Provincial Education Department (Grant No. 21JK0946), Beijing National Laboratory for Condensed Matter Physics (Grant No. 202162100001), and Double First-Class University Construction Project of Northwest University are acknowledged.

	\begin{appendix}
    \numberwithin{equation}{section}

	\section{\texorpdfstring{The second fusion branch}{The second fusion branch}}\label{APP:Fusion}
	
	Let us introduce the second fusion branch of $R$-matrix in Section \ref{Sec:second:fusion} detailedly.
	When $u=-\eta$, the $R$-matrix in (\ref{a}) becomes
	\begin{equation}\label{pro3}
		R_{1,2}(-\eta)=-2\eta{P}_{1,2}^{(-)}=-2\eta(1-{P}_{1,2}^{(+)}),
	\end{equation}
	where \({P}_{1,2}^{(-)}\) is a 2-dimensional supersymmetric projector with the following form
	\begin{align}
	&{P}_{1,2}^{(-)}=\sum_{i=1}^2|\bar{\psi}_i\rangle\langle \bar{\psi}_i|,\qquad {P}_{1,2}^{(-)}={P}_{2,1}^{(-)},\label{ju1}\\
	&|\bar{\psi}_1\rangle=\frac1{\sqrt2}(|1,2\rangle-|2,1\rangle),\quad|\bar{\psi}_2\rangle=|2,2\rangle.\label{A12}
	\end{align}
	The corresponding parities are
	\[p(\bar{\psi}_1)=1,\quad p(\bar{\psi}_2)=0.\]
	The operator \({P}_{1,2}^{(-)}\) projects the 4-dimensional product space $V_{1}\otimes_{s} V_{2}$ into a new 2-dimensional space spanned by \(\{\vert \bar{\psi}_i\rangle\vert i = 1,2\}\).

By fusing the $R$-matrix with this projector \({P}_{1,2}^{(-)}\), we can obtain the specific form of $R_{\bar{1}^{\prime},n}(u)$ defined in (\ref{bbb}), which is
	\begin{align}
	 R_{\bar{1}^{\prime},n}(u)=\begin{pmatrix}
		u+\frac{3}{2}\eta & & \\
		& u-\frac{1}{2}\eta & -\sqrt{2} \eta \\
		& -\sqrt{2} \eta & u+\frac{1}{2}\eta \\
		& & & u-\frac{3}{2}\eta
	\end{pmatrix}.
\end{align}

	At the point of $u=\tfrac{3}{2}\eta $, the fused $R$-matrix $R_{\bar{1}^{\prime},2}(u)$ in (\ref{bbb}) degenerates into
	\begin{equation}\label{pro4}
	R_{\bar{1}^{\prime},2}(\tfrac{3}{2}\eta )=3\eta \mathcal{P}_{\bar{1}^{\prime},2}^{(+)},
	\end{equation}
	where $\mathcal{P}_{\bar{1}^{\prime},2}^{(+)}$ is a 2-dimensional supersymmetric projector with the form of
	\begin{equation}\label{fu3}
		\mathcal{P}_{\bar{1}^{\prime},2}^{(+)}=\sum_{i=1}^2|\tilde{\phi}_{i}\rangle\langle\tilde{\phi}_{i}|,
	\end{equation}
	and the corresponding vectors are
	\begin{equation}\label{A16}
		|\tilde{\phi}_1\rangle=|\bar{\psi}_1\rangle\otimes_{s}|1\rangle,\quad|\tilde{\phi}_2\rangle=\frac{1}{\sqrt{3}}(\sqrt{2}|\bar{\psi}_2\rangle\otimes_{s}|1\rangle-|\bar{\psi}_1\rangle\otimes_{s}|2\rangle).
	\end{equation}	
Here, the $|\bar{\psi}_1\rangle$ and $|\bar{\psi}_2\rangle$ are given in  Eq. (\ref{A12}). The parities read
	\[p(\tilde{\phi}_1)=1,\quad p(\tilde{\phi}_2)=0.\]
	%The operator $\mathcal{P}_{\bar{1}^{\prime},2}^{(+)}$ projects the 4-dimensional product space  $V_{\bar{1}^{\prime}} \otimes_{s} V_{{2}}$ into a 2-dimensional space spanned by $\{\vert \tilde{\phi}_i\rangle, i = 1,2\}$.

Similarly, we can get the specific form of  the $R_{\tilde{1}^\prime,n}(u)$ given in Eq. (\ref{g})
\begin{align}
 R_{\tilde{1}^{\prime},n}(u)=\begin{pmatrix}
		u + 2\eta & & \\
		& u - \eta & -\sqrt{3}\eta \\
		& -\sqrt{3}\eta & u + \eta \\
		& & & u - 2\eta
	\end{pmatrix}.\label{Fused_R4}
\end{align}
From Eqs. (\ref{Fused_R2}) and (\ref{Fused_R4}), we can easily see that $R_{\tilde{1},2}(u)$ given by (\ref{f}) and $R_{\tilde{1}^\prime,2}(u)$ given by (\ref{g}) are the same, i.e., Eq. (\ref{m}).

	\section{Grassmann Numbers}\label{APP;Grassmann}

	Grassmann numbers are the anticommuting algebraic variables that play a central role in supersymmetric models and integrable systems with $\mathbb{Z}_2$ grading. 
	The Grassmann algebra $CG_N$ is generated by $N$ generators $\mathcal{E}_{1}$, $\mathcal{E}_{2}$,$\cdots$, $\mathcal{E}_{N}$, where the generators satisfy the nilpotency condition
	\begin{equation}\label{AG1}
		\mathcal{E}_{i}^2 = 0,
	\end{equation}
	and the anticommutation relations 
	\begin{equation}
		\mathcal{E}_i \mathcal{E}_j = - \mathcal{E}_j \mathcal{E}_i.
	\end{equation}

	\section{Proof of Eq. (\ref{o1})}\label{APPB}

We know that the reflecting monodromy matrix $\hat{T}(u)$ in Eq. (\ref{Tt11}) and its fused analogues satisfy the graded RTT relations
	\begin{equation}\label{oo1}
	R_{\alpha,\beta}(u-v) \hat{T}_{\alpha}(u) \hat{T}_{\beta}(v) = \hat{T}_{\beta}(v) \hat{T}_{\alpha}(u) R_{\alpha,\beta}(u-v),
	\end{equation}
	where the indices $\alpha, \beta$ may label either the original spaces or the projected spaces.
	
	Because the (fused) $R$-matrices collapse to projectors at certain special values of the spectral parameter, the (fused) monodromy matrices $\hat{T}_\alpha(u)$ satisfy the following relations
	 \begin{equation}
				\begin{aligned}\label{jie1}
					&P^{(+)}_{1,2}\hat{T}_1 (u)\hat{T}_2 (u+\eta)P^{(+)}_{1,2}=\prod_{l=1}^N
					(u+\theta_l+\eta)\hat{T}_{\bar 1}(u+\tfrac{1}{2}\eta),\\
					&{P}_{1,2}^{(-)}\hat{T}_1(u)\hat{T}_2 (u-\eta){P}_{1,2}^{(-)}=\prod_{l=1}^N
					(u+\theta_l-\eta)\hat{T}_{\bar 1^\prime}(u-\tfrac{1}{2}\eta),\\
					&\mathbb{P}_{2,\bar{1}}^{(-)} \hat{T}_{2} (u+\eta)\hat{T}_{\bar{1}}(u-\tfrac{1}{2}\eta)\mathbb{P}_{2,\bar{1}}^{(-)}=\prod_{l=1}^N
					(u+\theta_l)\hat{T}_{\tilde 1}(u),\\
					&\mathcal{P}_{2,\bar{1}^{\prime}}^{(+)} \hat{T}_{2} (u-\eta)\hat{T}_{\bar{1}^\prime}(u+\tfrac{1}{2}\eta)\mathcal{P}_{2,\bar{1}^{\prime}}^{(+)}=\prod_{l=1}^N
					(u+\theta_l)\hat{T}_{\tilde 1^\prime }(u),
				\end{aligned}
		\end{equation}
	where the projectors $P_{1,2}^{(+)}$, $\mathbb{P}_{2,\bar{1}}^{(-)}$, ${P}_{{1},2}^{(-)}$ and $\mathcal{P}_{2,\bar{1}^{\prime}}^{(+)}$ are given by (\ref{e}),(\ref{fu2}),  (\ref{ju1}) and  (\ref{fu3}), respectively.

		We define the degenerate point of the $R$-matrix as $\delta$, at which we have $R_{\alpha,\beta}(\delta)=P^{(d)}_{\alpha,\beta}S_{\alpha,\beta}$,
		where $P^{(d)}_{\alpha,\beta}$
		is a $d$-dimensional projector and $S_{\alpha,\beta}$ is a constant matrix. Employing the property of the projector that $P^{(d)}_{\alpha,\beta}R_{\alpha,\beta}(\delta)=R_{\alpha,\beta}(\delta)$, the RTT
		relations (\ref{d}) and (\ref{RTT-1}) at the degenerate point give
		\begin{equation}\label{fa}
				T_{\alpha}(u)T_{\beta}(u+\delta)P_{\beta,\alpha}^{(d)}=P_{\beta,\alpha}^{(d)}T_{\alpha}(u)T_{\beta}(u+\delta)P_{\beta,\alpha}^{(d)}.
		\end{equation}

		Similarly, from the graded RTT relations (\ref{oo1}), we have
		\begin{equation}
		\hat{T}_{\alpha} (u)\hat{T}_{\beta} (u+\eta)P^{(d)}_{\alpha,\beta}=P^{(d)}_{\alpha,\beta}\hat{T}_{\alpha} (u)\hat{T}_{\beta} (u+\eta)P^{(d)}_{\alpha,\beta},
	\end{equation}
	
	Using the properties of projector, one can derive the following identities from Eq. (\ref{jie1})  
	%Considering the GYBE (\ref{b4}) at the points of $u=-\theta_j$, $v=\{-\theta_j+\eta,\,-\theta_j-\eta\}$, Eq.(\ref{hat-1}) at the points of $u=-\theta_j,\,v=\{-\theta_j-\tfrac{3}{2}\eta, -\theta_j+\tfrac{3}{2}\eta\}$ and using the properties of projector, we get
	\begin{equation}
				\begin{aligned}
					& \Hat{T}_1(-\theta_j)\hat{T}_2(-\theta_j+\eta)=P_{1,2}^{(+)}\hat{T}_1(-\theta_j)\hat{T}_2(-\theta_j+\eta),\\
					& \hat{T}_1(-\theta_j)\hat{T}_2(-\theta_j-\eta)={P}_{{1},2}^{(-)}\hat{T}_1(-\theta_j)\hat{T}_2(-\theta_j-\eta),\\
					& \hat{T}_2(-\theta_j)\hat{T}_{\bar{1}}(-\theta_j-\tfrac{3}{2}\eta)=\mathbb{P}_{2,\bar{1}}^{(-)}\hat{T}_2(-\theta_j)\hat{T}_{\bar{1}}(-\theta_j-\tfrac{3}{2}\eta),\\
					& \hat{T}_2(-\theta_j)\hat{T}_{\bar{1}^{\prime}}(-\theta_j+\tfrac{3}{2}\eta)=\mathcal{P}_{2,\bar{1}^{\prime}}^{(+)}\hat{T}_2(-\theta_j)\hat{T}_{\bar{1}^{\prime}}(-\theta_j+\tfrac{3}{2}\eta),
				\end{aligned}\label{i-111}
		\end{equation}
where $j=1,\ldots, N$.
 
We can combine Eq. (\ref{i}) for the monodromy matrices $T_\alpha(u)$ and Eq. (\ref{i-111}) for the reflecting monodromy matrices $\hat{T}_\alpha(u)$ and finally get the following equations
\begin{equation}
		\begin{aligned}
& t(u)t(u+\eta)=[\rho_2(2u+\eta)]^{-1}{\rm str}_{1,2}\{K_2^+(u+\eta)R_{1,2}(-2u-\eta)K_1^+(u)T_1(u)T_2(u+\eta)\\[3pt]	&\hspace{4mm}\times K_1^-(u)R_{2,1}(2u+\eta)K_2^-(u+\eta)\hat T_1(u)\hat T_2(u+\eta)\},	
\end{aligned}\label{fus-1}
\end{equation}
\begin{equation}
		\begin{aligned}
& t^{(1)}(u-\tfrac{1}{2}\eta)t(u+\eta)=[\rho_3(2u+\tfrac{1}{2}\eta)]^{-1}{\rm str}_{\bar 1,2}\{K_{\bar 1}^{+}(u-\tfrac{1}{2}\eta)R_{2,\bar 1}(-2u-\tfrac{1}{2}\eta)K_2^+(u+\eta)\\[3pt]
&\hspace{4mm}\times T_2(u+\eta)T_{\bar 1}(u-\tfrac{1}{2}\eta) K_2^{-}(u+\eta)R_{\bar 1,2}(2u+\tfrac{1}{2}\eta)K_{\bar 1}^{-}(u-\tfrac{1}{2}\eta)\hat T_2(u+\eta)\hat T_{\bar 1}(u-\tfrac{1}{2}\eta)\},
\end{aligned}\label{fus-2}
\end{equation}
\begin{equation}
		\begin{aligned}
& t^{(2)}(u+\tfrac{1}{2}\eta)t(u-\eta)=[\rho_4(2u-\tfrac{1}{2}\eta)]^{-1}{\rm str}_{\bar {1}^{\prime},2}\{K_{\bar {1}^{\prime}}^{+}(u+\tfrac{1}{2}\eta)R_{2,\bar {1}^{\prime}}(-2u+\tfrac{1}{2}\eta)K_2^+(u-\eta)\\[4pt]
&\hspace{4mm}\times T_2(u-\eta)T_{\bar {1}^{\prime}}(u+\tfrac{1}{2}\eta) K_2^{-}(u-\eta)R_{\bar {1}^{\prime},2}(2u-\tfrac{1}{2}\eta)K_{\bar {1}^{\prime}}^{-}(u+\tfrac{1}{2}\eta)\hat T_2(u-\eta)\hat T_{\bar {1}^{\prime}}(u+\tfrac{1}{2}\eta)\}.
\end{aligned}\label{fus-3}
\end{equation}

Substituting Eq. (\ref{i-11}), (\ref{f1})-(\ref{f2}) and (\ref{fa})-(\ref{i-111}) into Eq. (\ref{fus-1}) and letting $u=\pm\theta_j,\pm\theta_j-\eta$ respectively, we get the first two lines of Eq. (\ref{o1}); substituting Eq. (\ref{i-11}), (\ref{f1})-(\ref{f2}) and (\ref{fa})-(\ref{i-111}) into Eq. (\ref{fus-2}) and letting $u=\pm\theta_j-\eta$, we get the third line of Eq. (\ref{o1});
substituting Eq. (\ref{i-11}), (\ref{f1})-(\ref{f2}) and (\ref{fa})-(\ref{i-111}) into Eq. (\ref{fus-3}) and letting $u=\pm\theta_j+\eta$, we get the fourth line of Eq. (\ref{o1}).

%$\{u=\pm\theta_j,\, \pm\theta_j-\eta\}$, into Eq. (\ref{fus-1}), using the result of Eq. (\ref{fus-2})with $\{u=\pm\theta_j-\eta\}$, Eq. (\ref{fus-3}) with $\{u=\pm\theta_j+\eta\}$ and combined with the relation (\ref{jie2}), we obtain the operators product relations (\ref{o1}).
\section{Commutation relations}\label{app:D}
Some useful commutation relations used in Section \ref{section4} are 
\begin{align} 
\tilde{\mathscr{C}}(u)\tilde{\mathscr{A}}(v)
&= \frac{(u-v+\eta)(u+v)}{(u+v+\eta)(u-v)}\,\tilde{\mathscr{A}}(v)\tilde{\mathscr{C}}(u)\nonumber\\
&\quad 
-\frac{\eta}{u+v+\eta}\Big\{\tilde{\mathscr{D}}(u)\tilde{\mathscr{C}}(v)+\frac{u+v}{u-v}\tilde{\mathscr{A}}(u)\tilde{\mathscr{C}}(v)\Big\} ,\label{com-1}\\
\tilde{\mathscr{D}}(v)\tilde{\mathscr{C}}(u)
&= \frac{(u-v-\eta)(u+v)}{(u+v-\eta)(u-v)}\,\tilde{\mathscr{C}}(u)\tilde{\mathscr{D}}(v)\nonumber\\
&\quad+ \frac{\eta}{u+v-\eta}\Big\{\tilde{\mathscr{C}}(v)\tilde{\mathscr{A}}(u)+\frac{u+v}{u-v}\,\tilde{\mathscr{C}}(v)\tilde{\mathscr{D}}(u)\Big\},\\
\tilde{\mathscr{A}}(u)\tilde{\mathscr{A}}(v)
		&=\tilde{\mathscr{A}}(v)\tilde{\mathscr{A}}(u)+\frac{\eta}{u+v+\eta}\Big\{\tilde{\mathscr{B}}(v)\tilde{\mathscr{C}}(u)-\tilde{\mathscr{B}}(u)\tilde{\mathscr{C}}(v) \Big\},\\
		\tilde{\mathscr{D}}(u)\tilde{\mathscr{D}}(v)
		&=\tilde{\mathscr{D}}(v)\tilde{\mathscr{D}}(u)-\frac{\eta}{u+v-\eta}\Big\{\tilde{\mathscr{C}}(v)\tilde{\mathscr{B}}(u)-\tilde{\mathscr{C}}(u)\tilde{\mathscr{B}}(v)\},\\
		\tilde{\mathscr{D}}(u)\tilde{\mathscr{A}}(v)
		&=\tilde{\mathscr{A}}(v)\tilde{\mathscr{D}}(u)-\frac{\eta(u+v)}{(u-v)(u+v+\eta)}\Big\{\tilde{\mathscr{B}}(v)\tilde{\mathscr{C}}(u)-\tilde{\mathscr{B}}(u)\tilde{\mathscr{C}}(v)\Big\},\label{com-2}\\
		\tilde{\mathscr{B}}(u)\tilde{\mathscr{B}}(v)
		&=-\frac{u-v-\eta}{u-v+\eta}\,\tilde{\mathscr{B}}(v)\tilde{\mathscr{B}}(u),\\
		\tilde{\mathscr{C}}(u)\tilde{\mathscr{C}}(v)
		&=-\frac{u-v+\eta}{u-v-\eta}\,\tilde{\mathscr{C}}(v)\tilde{\mathscr{C}}(u).
\end{align}

\end{appendix}

%\bibliographystyle{SciPost_bibstyle}
%\bibliography{Ref_gl11}

\end{document}